\documentclass[draft,jgrga]{AGUTeX}

\usepackage{lineno}
\linenumbers*[1]

  \usepackage{graphicx}  
  \usepackage{amssymb}
\usepackage{amsmath}
\usepackage{url}

 \setkeys{Gin}{draft=false}
\newcommand{\abs}[1]{\left |#1\right |}

\newcommand{\paren}[1]{\left(#1\right)}
\newcommand{\bracket}[1]{\left[#1\right]}
\newcommand{\set}[1]{\left\{#1\right\}}

\newcommand{\mean}[1]{\langle#1\rangle}
\newcommand{\eps}{\epsilon}

\newcommand{\dt}{\partial_{t}}

\newcommand{\f}{\phi}

\newcommand{\calm}{\mathcal{M}}
\newcommand{\calM}{\mathcal{M}}
\newcommand{\calQ}{\mathcal{Q}}
\newcommand{\calR}{\mathcal{R}}

\newcommand{\calP}{\mathcal{P}}

\newcommand{\calV}{\mathcal{V}}

\newcommand{\bx}{\mathbf{x}}
\newcommand{\by}{\mathbf{y}}
\newcommand{\bv}{\mathbf{v}}
\newcommand{\bV}{\mathbf{V}}

\newcommand{\bg}{\mathbf{g}}
\newcommand{\bk}{\mathbf{k}}

\newcommand{\bn}{\mathbf{n}}

\newcommand{\meanrho}{\overline{\rho}}

\newcommand{\fzero}{{f(0)}}
\newcommand{\fone}{{f(1)}}

\newcommand{\szero}{{s(0)}}
\newcommand{\sone}{{s(1)}}
\newcommand{\eff}{{\textrm{eff.}}}

\authorrunninghead{SIMPSON ET AL.}

\titlerunninghead{A MULTISCALE MODEL OF PARTIAL MELTS}

\authoraddr{G. Simpson,
Department of Mathematics, University of Toronto, Toronto, ON M5S 2E4, Canada.
(simpson@math.toronto.edu)}

\authoraddr{M. Spiegelman,
Department of Applied Physics and Applied Mathematics, Columbia University, New York, NY 10027, USA.
Lamont-Doherty Earth Observatory, Palisades, NY 10964, USA.
(mspieg@ldeo.columbia.edu)}

\authoraddr{M. I. Weinstein,
Department of Applied Physics and Applied Mathematics, Columbia University, 200 Mudd, New York, NY 10027, USA.
(miw2103@columbia.edu)}

\begin{document}

%
%

\title{A Multiscale Model of Partial Melts 1: Effective Equations}
%

%
%


\author{G. Simpson}
\affil{Department of Mathematics, University of Toronto, Toronto, ON M5S 2E4, Canada.}

\author{M. Spiegelman}
\affil{Department of Applied Physics and Applied Mathematics,
Columbia University, New York, New York, USA.  Lamont-Doherty Earth Observatory, Palisades, NY 10964, USA.}

\author{M. I. Weinstein}
\affil{Department of Applied Physics and Applied Mathematics,
Columbia University, New York, New York, USA.}



%
%
%

%
%


\begin{abstract}
  Developing accurate and tractable mathematical models for partially
  molten systems is critical for understanding the dynamics of
  magmatic plate boundaries as well as the geochemical evolution of
  the planet.  Because these systems include interacting fluid and
  solid phases, developing such models can be challenging.  The
  composite material of melt and solid may have emergent properties, such as
  permeability and compressibility, that are absent in the
  phase alone.  Previous work by several authors have used
  multiphase flow theory to derive macroscopic equations based on
  conservation principals and assumptions about interphase forces and
  interactions.  Here we present a complementary approach using
  homogenization, a multiple scale theory.  Our point of
  departure is a model of the microstructure, assumed to possess an
  arbitrary, but periodic, microscopic geometry of interpenetrating
  melt and matrix.  At this scale, incompressible Stokes flow is
  assumed to govern both phases, with appropriate interface
  conditions.
  
  Homogenization systematically leads to macroscopic equations for the
  melt and matrix velocities, as well as the bulk parameters,
  permeability and bulk viscosity, without requiring ad-hoc closures
  for interphase forces. We show that homogenization can lead to a
  range of macroscopic models depending on the relative contrast in
  melt and solid properties such as viscosity or velocity. In
  particular, we identify a regime that is in good agreement with
  previous formulations, without including their attendant
  assumptions. Thus this work serves as independent verification of
  these models.  In addition, homogenization provides a consistent
  machinery for computing consistent macroscopic constitutive
  relations such as permeability and bulk viscosity that are
  consistent with a given microstructure. These relations are explored
  numerically in a companion paper.
\end{abstract}


%
%

%

\begin{article}

%
%

\section{Introduction}
  
Developing quantitative models of partially molten regions in the
Earth is critical for understanding the dynamics of magmatic plate
boundaries such as mid-ocean ridges and subduction zones, as well as
for providing a better integration of geochemistry and geophysics.
Beginning in the mid 1980's there have been multiple derivations of
macroscopic equations for magma dynamics that describe the flow of a
low-viscosity fluid in a viscously deformable, permeable solid matrix
\cite[]{mckenzie1984gac, scott1984ms, scott1986map, fowler1985mmm,
  fowler1989gac, Spiegelman1993a, stevenson1991mfr,
  bercovici2001a,bercovici2001c,ricard2001b, hiermajumder2006rgb,
  bercovici2005tpg, bercovici2003etp, ricard2003tpd, ricard2007pmc}.
The details and specific processes included, vary slightly among these model
systems but all are derived using the methods of multiphase flow
\cite[e.g.][]{drew1971aet,drew1971afe,drew1983mmp}.  

Multiphase flow techniques are well formulated in many texts, including \cite{drew1999tmf,brennen2005fmf}.
Typically, the two-phase medium is examined at a macroscopic scale, much larger
than the pore or grain scale, and one attempts to develop effective media
equations based on conservation of mass, momentum, and energy.  This approach
is reasonably straightforward and has proven useful in applications, notably dilute disperse flows.
However, they have two fundamental sources of uncertainty. First, an
appropriate ``interphase force'' must be posited.  There
is often a tremendous range of mathematically valid choices for this force with
little to constrain it beyond physical intuition and experimental validation.
Second, the macroscopic equations derived for the partial melt problem include 
critical constitutive relations, such as permeability, bulk
viscosity or effective shear viscosity.   These should depend on the
microscopic distribution of melt and matrix, information that is often lost in the 
multiphase flow approach. Multiphase flow does not naturally determine these relationships. 
As with the interphase force, these closures require estimates from scalings, 
numerical simulations, and experiments.


In this \nocite{simpson08a} 
 and a companion paper \cite[]{simpson08b}, we present a
complementary method for deriving effective macroscopic equations
using the methods of homogenization theory. Rather
than assume macroscopic equations and then seek closures for
constitutive relations, we assume microscopic equations and derive the
macroscopic equations.  This is done by a multiple scale expansion,
which encodes both fine and coarse length scales into the field
variables.  As in all multiple scale methods, the equations are
matched at each order of the small parameter and solved successively. 
For a  useful introduction to homogenization with applications see
\cite{hornung1997hap, torquato2002rhm}.  More rigorous mathematical
treatments are presented in \cite{bensoussan1978aap,
  sanchezpalencia1980nhm, cioranescu1999ih, checkin2007hma,
  pavliotis2008mma}.   Homogenization has been used extensively for
flow in rigid and elastic porous media, but we believe this is the first application to
the magma dynamics problem which permits viscous deformation of the matrix. 

This strategy has several advantages with respect to multiphase flow
methods.  In particular, there is no under-constrained interphase
force, as these effects are described precisely by boundary conditions between
the phases at the micro scale. Second, and perhaps more importantly,
it provides a mechanism for computing consistent macroscopic
constitutive relations for a given microscopic geometry.  For the magma dynamics 
problem, it yields a collection of auxiliary ``cell problems'' whose solutions determine the bulk
viscosity, shear viscosity, and permeability of the medium consistent
with the micro-structure.  We emphasize that these macroscopic
effective quantities are not volume averages of microscopic
quantities.  Indeed, permeability and bulk viscosity are undefined at
the grain scale, but they appear as macroscopic properties through
homogenization. More generally, homogenization can extract tensor valued
permeabilities and shear viscosities for anisotropic media.  The methods presented 
are also adaptable to other fine scale rheologies and physics.

In this work, we specifically consider the simplest case of
homogenization of the momentum equations for two coupled Stokes
problems involving a high viscosity phase (the solid matrix) and a low
viscosity fluid. This work is adapted from studies of sintering and
partially molten metal alloys,
\cite[]{auriault1992mhc,geindreau1999ivb}, which in turn is based  on earlier work in poro-elastic media  \cite[e.g.,][]{auriault1987ndp,auriault1991hme,auriault1992dpm,mei1989mhp,mei1996sah, auriault2002swf}.  For clarity,
we only consider linear viscous behavior for the solid, as
this may be appropriate for the diffusion creep regime \cite[e.g.,
][]{hirth1995ecd1}.  This assumption considerably simplifies the
analysis. Extensions to power-law materials are discussed in \cite{geindreau1999ivb}.

We demonstrate that, depending on the choice of scalings, we can derive
homogenized macroscopic equations for three different regimes, and
identify a particular regime that is consistent with existing and
commonly used formulations such as \cite{mckenzie1984gac} and
\cite{bercovici2003etp}.  This provides independent validation of
these other systems.   We also discuss the strengths and
weaknesses of homogenization and identifies some open questions.  We
 recognize that the derivation is somewhat technical but
have attempted to make the overall approach as accessible as possible
with the hope that other researchers will extend these methods to
 related problems (e.g., including surface energies and more
complex rheologies).

The second paper is more practical and provides specific
computation of cell-problems to calculate consistent permeabilities,
bulk-viscosities and effective shear viscosities for several
simplified pore geometries.  In particular we provide a
derivation for the bulk viscosity and demonstrate that, for a purely
mechanical coupling of phases, it should scale inversely with the
porosity; a relationship we conjecture is insensitive to the specifics
of the microscopic geometry.  Such an inverse relationship has been
suggested before \cite[e.g.\ ][]{schmeling2000pma,bercovici2003etp};
however, this is the first rigorous derivation from the microscale.
Further implications of these results are discussed in the second
paper.

\cite{simpson08a, simpson08b} are based on the PhD. thesis of G.~Simpson, \cite{simpson08t}.

\section{Problem Description}
\label{sec:problem-description}

\subsection{Macroscopic and Microscopic Domains}
\label{sec:domains}

\begin{figure}
\noindent\includegraphics[width= 15pc]{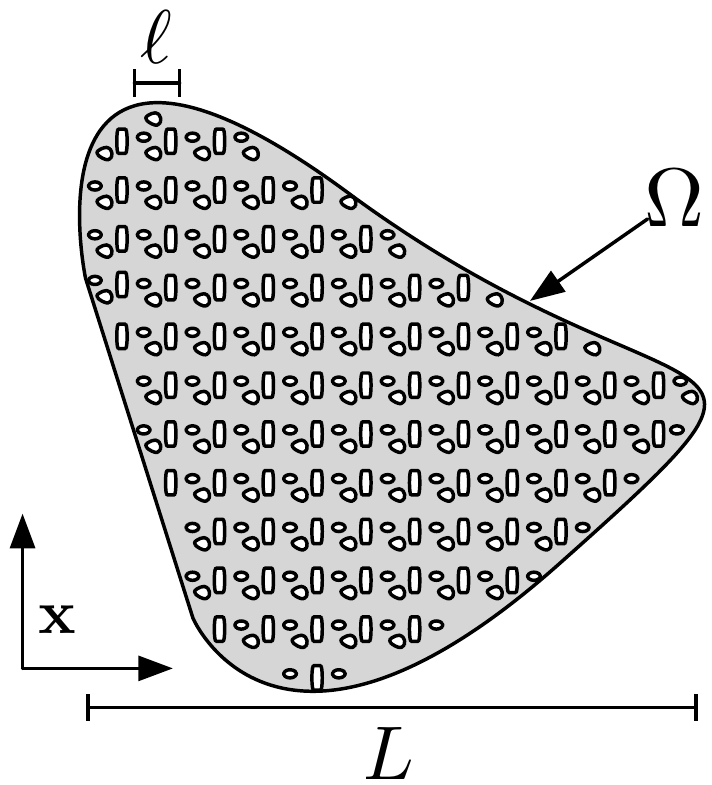}
\caption{The macroscopic domain $\Omega$.  The matrix occupies the
  gray region while the  melt occupies the white inclusions.}
\label{fig:homog_domain}
\end{figure}

To begin the upscaling procedure, we must describe the spatial domains
occupied by each phase. We denote with symbol $\Omega$ the total
macroscopic region of interest, containing both the melt and the
matrix, with a characteristic length scale $L$ which might be an
observed macroscopic characteristic wavelength (e.g. 1 m--10 km). 
Initially, we assume that within $\Omega$ the matrix has a periodic
microstructure.   A two-dimensional analog is pictured in Figure
\ref{fig:homog_domain}.  $\Omega$, the bounded gray region, is tiled
with a fluid filled pore network of period $\ell$.  $\ell$ is a
representative measure of length scale of the grains or pore distribution, such as a statistical moment of the grain size distribution,  and is much smaller than $L$.   Notation for the domains is given in Table \ref{table:homog_domains}.

  \begin{table}
  \centering
  \caption{Notation for domains in homogenization model.}
  \label{table:homog_domains}
  \begin{tabular}{rl}
  \hline\hline
  Symbol & Meaning \\
  \hline
  $\gamma$& Interface between melt and matrix within cell $Y$\\
  $\Gamma$&Total macroscopic interface between melt and matrix\\
  $\Omega$ &Total macroscopic space occupied by both melt and matrix\\
  $\Omega_f$ &Portion of macroscopic space occupied by melt\\
  $\Omega_s$ &Portion macroscopic space occupied by matrix\\
  $Y$ & The unit cell\\
  $Y_f$ & Portion of unit cell occupied by melt\\
    $Y_s$ & Portion of unit cell occupied by matrix\\
    \hline
  \end{tabular}
\end{table}

We form the first important  dimensionless parameter, $\eps$,
\begin{equation}
\label{eq:homog_eps}
\boxed{\eps \equiv \frac{\ell}{L}}
\end{equation}
$\eps$ will play two important roles in what follows.  First, all other dimensionless numbers and parameters will be expressed in powers of $\eps$.  Second, we will expand the dependent variables in powers in $\eps$ as in
\begin{equation}
\label{eq:pseries1}
\Phi = \Phi^{(0)} + \eps \Phi^{(1)} + \eps^2 \Phi^{(2)} + \ldots
\end{equation}

\begin{figure}
\noindent\includegraphics[width=20pc]{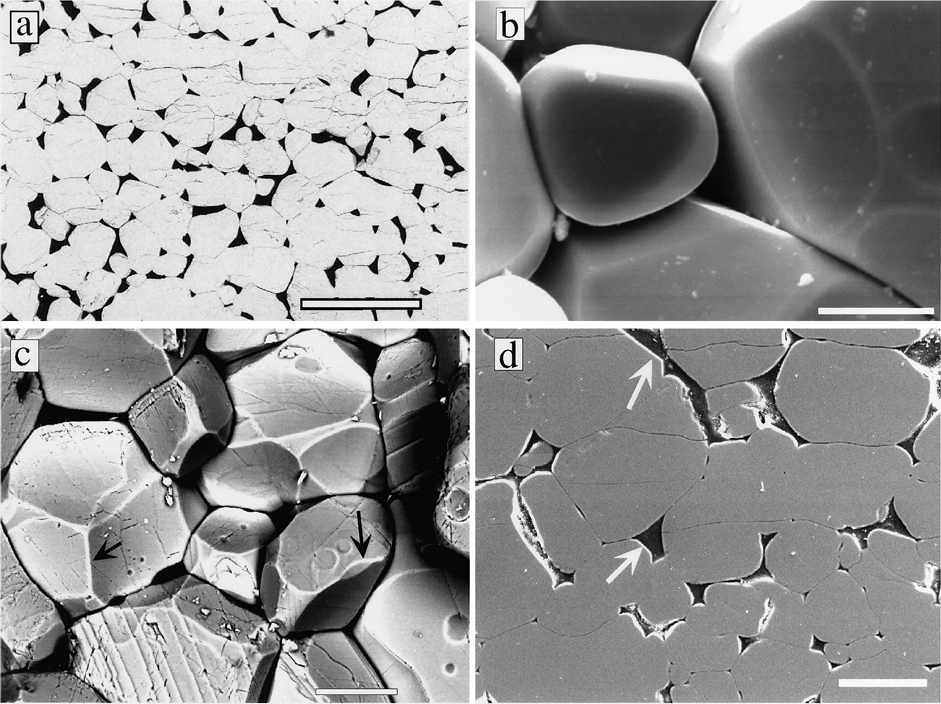}
\caption{SEM images of synthetic quartzites and marbles from Figure 5
  of \cite{wark1998gsp}. Similar microstructures are seen in olivine
  basalt aggregates \cite[e.g.][]{hirth1995ecd1,hirth1995ecd2}}
\label{fig:wark_pores}
\end{figure}

Of course, real partially molten rocks are not a periodic medium.
Pore structures similar to those expected in peridotite appear in
Figure \ref{fig:wark_pores}.  Since it is crystalline, there is some
regularity, but it is closer to a \emph{random} medium.  While only
the periodic case is treated in this work, the random one is of
interest and is also amenable to homogenization, \cite{torquato2002rhm}.

We divide our domain $\Omega$ from Figure \ref{fig:homog_domain} into three subregions:
\begin{align*}
\Omega_f - &\text{The fluid portion of $\Omega$.}\\
\Omega_s - &\text{The solid portion of $\Omega$.}\\
\Gamma - &\text{The interface between fluid and solid in $\Omega$.}
\end{align*}
We shall write equations for the melt in $\Omega_f$, equations for the matrix in $\Omega_s$, and boundary conditions between the two along $\Gamma$.

\begin{figure}
\noindent\includegraphics[width=12pc]{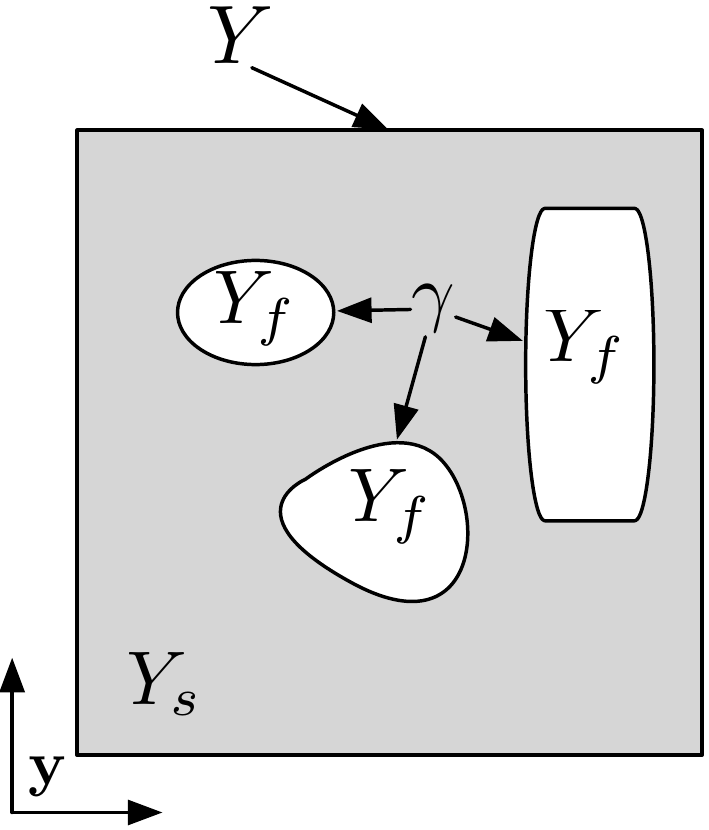}
\caption{The cell domain, $Y$, divided into fluid and solid regions, $Y_f$ and $Y_s$.  The two phases meet on interface $\Gamma$.}
\label{fig:cell_domain}
\end{figure}

\begin{figure}
\noindent\includegraphics{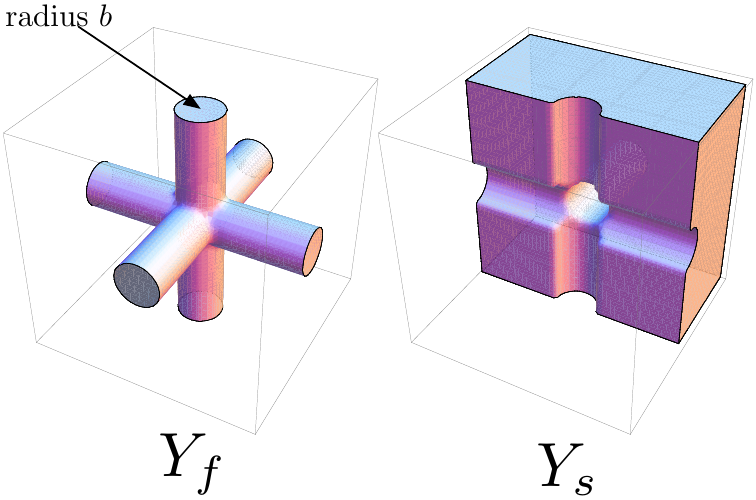}
\caption{A cell geometry in which both the fluid region, $Y_f$, and the solid region, $Y_s$, are topologically connected.}
\label{fig:tube_geometry}
\end{figure}

We now introduce the notion of a cell.  The cell, appearing in 
Figure \ref{fig:cell_domain} and denoted with the symbol $Y$, 
is a scaled, dimensionless, copy of the periodic microstructure 
of Figure \ref{fig:homog_domain}.  This is divided into a fluid 
region, $Y_f$, a solid region, $Y_s$, and an interface, $\gamma$.  
A simple three-dimensional example of such a cell appears 
is displayed in Figure \ref{fig:tube_geometry}.   The cell should 
be interpreted as a scaled representative elementary volume 
of the grain scale.  It may be a single grain or a small ensemble 
of grains.


Although the connectedness of both phases is an important property,
the particular microstructure of $Y$ does not play a significant role
in the form of macroscopic equations.  Cell geometry does
determine the magnitudes and forms of the constitutive relations 
appearing in the equations.  This is discussed in the companion paper.  


\subsection{Grain Scale Equations}
\label{sec:grainscale}

At the microscale, we assume both phases are incompressible,
 linearly viscous, isotropic fluids.  The rheology for each phase is:
\begin{equation}
\boxed{\sigma = - p I + 2\mu e\paren{\bv}}
\end{equation}
where $e(\bv)$ is the strain-rate tensor,
\begin{equation}
\boxed{e(\bv) = \frac{1}{2}\paren{\nabla\bv + (\nabla\bv)^T}}
\end{equation}
Components may be accessed by index notation:
\begin{align*}
e_{ij}(\bv) &= \frac{1}{2}\paren{\frac{\partial v_i}{\partial X_j}+ \frac{\partial v_j}{\partial X_i}}\\
\sigma_{ij} &= -p\delta_{ij} + 2\mu e_{ij}(\bv)
\end{align*}
The variable $\mathbf{X}$ appearing in these expressions 
denotes the dimensional spatial variable.
The stress in $\Omega_s$ for the solid phase is $\sigma^{s}$, 
with pressure $p^{s}$ and velocity $\bv^{s}$.  Similarly, the 
fluid has stress $\sigma^{f}$, pressure $p^{f}$, and velocity 
$\bv^{f}$ in $\Omega_f$.   The notation for the fields is 
given in Table \ref{table:homog_fields}.

  \begin{table}
  \centering
  \caption{Notation for fields in homogenization model.}
  \label{table:homog_fields}
  \begin{tabular}{rl}
  \hline\hline
  Symbol & Meaning \\
  \hline
  $e(\bv)$ & Strain rate tensor, $e(\bv) = \frac{1}{2}(\nabla \bv + (\nabla \bv)^T)$\\
  $\phi$& Volume fraction of melt, $\phi \equiv \int_{Y_f} d\by$  \\
  $\mathbf{g}$& $-g \mathbf{z}$ \\
  $\mathbf{g}^f$& $\rho_f \mathbf{g}$ \\
  $\mathbf{g}^s$& $\rho_s \mathbf{g}$ \\
 $p^f$ & Melt pressure\\
  $p^{f(j)}$ & Melt pressure at order $j$ in the series expansion\\
  $p^s$ & Melt pressure\\
  $p^{s(j)}$ & Matrix pressure at order $j$ in the series expansion\\
  $\sigma^f$&Melt Stress Tensor \\
  $\sigma^{f(j)}$&Melt Stress Tensor at order $j$ in the series expansion\\
  $\sigma^s$&Matrix Stress Tensor\\
  $\sigma^{s(j)}$&Matrix Stress Tensor at order $j$ in the series expansion\\
  $\mathbf{v}^f$& Melt velocity\\
  $\mathbf{v}^{f(j)}$& Melt velocity at order $j$ in the series expansion\\
  $\mathbf{v}^s$&Matrix Velocity\\
  $\mathbf{v}^{s(j)}$& Matrix velocity at order $j$ in the series expansion\\
    \hline
  \end{tabular}
\end{table}

  \begin{table}
  \centering
  \caption{Notation and measurements for models of partial melts.}
  \label{table:measurements}
  \begin{tabular}{rll}
  \hline\hline
  Symbol & Meaning & Value\\
  \hline
  $\phi$& Volume Fraction of Melt & $.01 \% $-- $10\%$\\
  $g$&Gravity& 9.8 $\mathrm{m/s^2}$\\
   $\ell$& Grain Length Scale &1 --10 mm\\
     $L$& Macroscopic Length Scale&1 m - 10 km\\
  $\mu_f$&Melt Shear Viscosity&$1$--$10$ Pa s\\
  $\mu_s$&Matrix Shear Viscosity&$10^{15}$--$10^{21}$ Pa s\\
  $\mathrm{Re}^f_\ell$& Reynolds Number of Melt&$10^{-8}$--$10^{-5}$\\
  $\mathrm{Re}^s_\ell$& Reynolds Number of Matrix& $10^{-30}$--$10^{-22}$\\
  $\rho_f$&Melt Density &$2800$ kg/m$^3$\\
  $\rho_s$&Matrix Density& $3300$ kg/m$^3$\\
   $V^f$& Characteristic Melt Velocity&1 -- 10 m/yr\\
 $V^s$& Characteristic Matrix Velocity&1 -- 10 cm/yr\\
    \hline
  \end{tabular}
\end{table}

At the pore scale, the Reynolds number is small; using the the 
values of Table \ref{table:measurements}, the value in the melt 
is $\lesssim O(10^{-5})$ and as low as $O(10^{-30})$ in the 
matrix.  Therefore, we will omit inertial terms in the conservation 
of momentum equations.  Each phase satisfies the Stokes 
equations at the grain scale; the divergence of the stress of 
each phase balances the body forces.  As we are interested in 
buoyancy driven flow, the forces $\bg^{s}\equiv  -\rho_s g \mathbf{e}_3$ 
and $\bg^{f}\equiv -\rho_f g \mathbf{e}_3$ are included.  The equations are:
\begin{subequations}
\begin{align}
\label{eq:force}
\nabla \cdot \sigma^{f}+\bg^{f} &= 0\quad\text{in $\Omega_{f} $}& \quad \nabla \cdot \sigma^{s}+\bg^{s} &= 0\quad\text{in $\Omega_{s} $}\\
\label{eq:inc}
\nabla \cdot\bv^{f}&= 0\quad\text{in $\Omega_{f} $}&\quad \nabla \cdot\bv^{s}&= 0\quad\text{in $\Omega_{s} $}
\end{align}
\end{subequations}
Conditions at the interface between fluid and solid, $\Gamma$, 
are still needed.  As both are viscous, we posit continuity of 
velocity and normal stress:
\begin{subequations}
\begin{align}
\label{eq:stress_bc}
\bv^{s}&=\bv^{f},\quad\text{on $\Gamma$}\\
\label{eq:velocity_bc}
\sigma^{s}\cdot \bn &=\sigma^{f}\cdot \bn,\quad\text{on $\Gamma$}
\end{align}
\end{subequations}
A Boussinesq approximation has been made by taking the velocities to
be continuous as opposed to the momenta.  These equations are {exact}
in the sense that, subject to boundary conditions on the exterior of
$\Omega$, solving them would provide a full description of the
behavior of the two-phase medium (although it would be impractical to
solve such a system at the macroscopic scale of interest).

\subsection{Scalings}
\label{sec:scalings}
Our effective equations emerge from multiple scale expansions of the dependent variables.  The dimensional spatial variable $\mathbf{X}$ specifies a position within either $\Omega_s$ or $\Omega_f$.  We introduce two dimensionless spatial scales, $\by$, the ``fast" spatial scale, and $\bx$, the ``slow" spatial scale.  These relate to $\mathbf{X}$, and one another, as:
\begin{subequations}
\begin{align}
\mathbf{y}&\equiv \mathrm{\mathbf{X}}/\ell\\
\mathbf{x}&\equiv \mathrm{\mathbf{X}}/L = \eps \mathbf{y}
\end{align}
\end{subequations}

The expansion in \eqref{eq:pseries1} is now made more precise.  All variables are assumed, initially, to have both fast and slow scale dependence:
\begin{equation}
\label{eq:variable_expansion}
\Phi(\by)=\Phi ^{(0)}(\eps \by,\by )+\eps\Phi ^{(1)}(\eps \by,\by )+ \eps^2\Phi ^{(2)}(\eps \by,\by)+\ldots
\end{equation}
Such an expansion captures  grain scale detail in the second 
argument, while permitting slow, macroscopic variations in the 
first argument.  As we take our domain to be periodically tiled 
with scaled copies of the cell $Y$, we assume $\Phi^{(j)}(\bx,\by)$ 
is $\by$-periodic at all orders of $j$.  We seek equations that 
are {only} functions of $\bx$; these will be the effective 
macroscopic equations.

Before making series expansions in powers of $\eps$, the 
equations must be scaled appropriately.  In addition to $\eps$, 
there are several other important dimensionless numbers.  
As motivation, let $P^s$, $P^f$, $V^s$, $V^f$ be characteristic 
pressures and velocities for the solid and fluid phases.  We write:
\begin{align}
p^{f} & = P^{f} \tilde{p}^{f} & p^{s} & = P^{s} \tilde{p}^{s}\\
\bv^{f} &= V^{f} \tilde{\bv}^{f}&\bv^{s} &= V^{s} \tilde{\bv}^{s}
\end{align}
Tildes reflect that the variables are now dimensionless and $O(1)$.  
Using these definitions, we non-dimensionalize \eqref{eq:force}.   
We are free to scale the equations to either the slow or fast length 
scale.  In this work, we scale to the $\ell$ length, though this does 
not affect the results.  On the length scale $\ell$, the force equations are:
\begin{equation}
\label{eq:scaled_force}
\begin{split}
\nabla_y \cdot \bracket{-\tilde{p}^{f} I +\paren{ \frac{\mu_{f} V^{f}}{P^{f} \ell}}2\tilde{\mu}_{f} e_y(\tilde{\bv}^{f}) }+\frac{\rho_{f}g\ell}{P^{f}}\tilde{\bg}^{f}&=0\\
\nabla_y \cdot \bracket{-\tilde{p}^{s} I +\paren{ \frac{\mu_{s} V^{s}}{P^{s} \ell}}2\tilde{\mu}_{s} e_y(\tilde{\bv}^{s}) }+\frac{\rho_{s}g\ell}{P^{s}}\tilde{\bg}^{s}&=0
\end{split}
\end{equation}
where $e_y$ denotes the strain-rate tensor with velocity gradients
taken with respect to the fast length scale $y$. 
This motivates defining four more dimensionless numbers:
\begin{align}
\label{eq:qparam_fast}
\mathcal{Q}^{f}_\ell & \equiv  \frac{\mu_{f} V^{f} }{P^{f} \ell}&\mathcal{Q}^{s}_\ell & \equiv  \frac{\mu_{s} V^{s} }{P^{s} \ell}\\
\label{eq:rparam_fast}
\mathcal{R}^{f}_\ell& \equiv \frac{\rho_{f}g\ell}{P^{f}}&\mathcal{R}^{s}_\ell& \equiv \frac{\rho_{s}g\ell}{P^{s}}
\end{align}
The $\calQ$'s measure the relative magnitudes of the viscous forces 
and the pressure gradients, while the $\calR$'s measure the relative 
magnitudes of the body forces and the pressure gradients.  The 
$\tilde{\mu}$'s and $\tilde{\bg}$'s remain in the equations as 
$O(1)$ constants.  Three other important parameters are the ratio of 
the viscosities of the two phases, the ratio of the velocities of the two 
phases, and the ratio of the pressures of the phases:
\begin{align}
\label{eq:params4}
\calm & \equiv{\frac{\mu_f}{\mu_s}}\\
\label{eq:params3}
\mathcal{V} & \equiv{\frac{V^f}{V^s}}\\
\label{eq:params5}
\mathcal{P} & \equiv{\frac{P^f}{P^s}}
\end{align}
A full list of dimensionless numbers is given in Table \ref{table:dimensionless_numbers}.

  \begin{table}
  \centering
  \caption{Dimensionless numbers for homogenization model.}
  \label{table:dimensionless_numbers}
  \begin{tabular}{rp{8cm}l}
  \hline\hline
  Symbol & Meaning & Estimate \\
  \hline
  $\eps$ & Length scale ratio, $\eps = \ell /L$ &  $O(10^{-7}- 10^{-2})$ \\
  $\calM$& Viscosity ratio,  $\calM =\mu_f/ \mu_s$ & $O(10^{-21} - 10^{-14})$ \\
$\calP$& Pressure ratio, $\calP = P^f / P^s$ & $O(\eps^0)$ \\
  $\calQ^f_\ell$ & Ratio of viscous force to pressure gradient in melt, $\calQ^f_\ell = (\mu_f V^f) / (P^f \ell)$& $O(\eps^9 - \eps^0)$\\
  $\calQ^s_\ell$ & Ratio of viscous force to pressure gradient in matrix, $\calQ^s_\ell = (\mu_s V^s) / (P^s \ell)$&$O(\eps^{-1})$\\
  $\calQ^s_L$ & Ratio of viscous force to pressure gradient in matrix, $\calQ^s_L = (\mu_s V^s) / (P^s L)$&$O(\eps^0)$\\
  $\calR^f_\ell$ & Ratio of buoyancy force to pressure gradient in melt, $\calR^f_\ell = (\rho_f g \ell) / P^f $&$O(\eps^1)$\\
  $\calR^s_\ell$ & Ratio of buoyancy force to pressure gradient in matrix, $\calR^s_\ell = (\rho_s g \ell) /P^s$&$O(\eps^1)$\\
  $\calR^s_L$ & Buoyancy force to pressure gradient ratio in matrix, $\calR^s_L = (\rho_s g L) /P^s$&$O(\eps^0)$\\
  $\calV$ & Velocity ratio, $\calV = V^f / V^s$ & $O(10^{-7}- 10^{-2})$\\
    \hline
  \end{tabular}
\end{table}

Starting with $\eps$, $\mathcal{V}$, and $\mathcal{M}$, we estimate 
these parameters with the data in Table \ref{table:measurements}:
\begin{align}
\calm &=  O(10^{-21} - 10^{-14})\\
\calV& = O(10^{1} -10^{3})\\
\eps &= O(10^{-7}- 10^{-2})
\end{align}
For these values, $\calm \ll \eps \ll \calV$. 
Since we expand the equations in powers of $\eps$, we relate all 
quantities to $\eps$. A quantity $Q$ is said to be $O(\eps^p)$ if
\begin{equation}
\eps^{p+1}\ll Q \ll \eps^{p-1}
\end{equation}
In terms of $\eps$, $\calM$ and $\calV$ are approximately:
\begin{align}
\calM & = O(\epsilon^{11} - \epsilon ^{2})\\
\calV & =  O(\epsilon ^{0} - \epsilon ^{-2})
\end{align}
We emphasize that this power of $\eps$ scale is less precise than a
power of 10 scale. 
%
For example, $\calV$ might be $O(10^1)$, but if
$\eps=O(10^{-4})$, we would say $\calV = O(\eps^0=1)$ since $\calV \ll
\eps^{-1}$.  Indeed, in one of the scaling regimes we consider, $\calV
= O(\eps^0=1)$.

To estimate the other parameters, we need estimates of the
characteristic pressures.  To do this, we first consider the forces on the
matrix.  At the macroscopic scale, the melt is $O(1\%)$ of the
medium's  volume.  We thus argue that on this scale, the matrix is
``close" to satisfying the Stokes equations; the pressure gradient,
viscous forces, and gravity balance one another.  On the large length
scale $L$, the dimensionless form of \eqref{eq:force} is
\begin{equation}
\nabla_x \cdot \bracket{-\tilde{p}^s I +\paren{ \frac{\mu_s V^s}{P^s L}}2\tilde{\mu}_s e_x(\tilde{\bv}^s) }+ \frac{\rho_sgL}{P^s}\tilde{\bg}^s=0.
\end{equation}
Similar to Equation \eqref{eq:scaled_force}, we define
\begin{align}
\label{eq:qparam-solid-slow}
\calQ^s_L &\equiv \frac{\mu_s V^s}{ P^s L}\\
\label{eq:rparam-solid-slow}
\calR^s_L &\equiv \frac{\rho_sg L}{P^s}
\end{align}
For the terms to be in balance, $\calQ^s_L = O(1)$ and $\calR^s_L = O(1)$.  Using (\ref{eq:qparam-solid-slow}--\ref{eq:rparam-solid-slow}) and the definition of $\eps$, the fast length scale parameters are:
\begin{align}
\calQ^s_\ell &= O\paren{\eps^{-1}}\\
\calR^s_ \ell&= O\paren{\eps ^{1}}
\end{align}

In the absence of direct pressure measurements, we assume the pressures are the same order,
\begin{equation}
\label{eq:pressure_scaling}
\calP=O(1=\eps^0).
\end{equation}
An argument for this is given in \cite{drew1983mmp}.  Briefly, since the velocities of interest are far less than the speed of sound, it would be difficult to support large pressure gradients across the phases without surface tension, a mechanism we do not include.  We make an additional assumption that there are $O(1)$ non-hydrostatic pressures in both phases; if $p = p_\textrm{hydro}+ p_\textrm{non-hydro}$ then
\begin{equation}
\label{eq:non_hydro}
\abs{\frac{p}{p_\textrm{non-hydro}}} = O(1=\eps^0).
\end{equation}

In the fluid, since $\rho_f/\rho_s = O(1)$, a consequence of $\calP=O(1)$ is
\begin{equation}
\label{eq:fluid_pressure_gravity}
\calR^f_ \ell = \frac{\rho_f g \ell}{P^f} = \frac{\rho_s g \ell}{P^s}\frac{\rho_f}{\rho_s}\frac{P^s}{P^f}  = \calR^s_\ell \frac{\rho_f}{\rho_s}  \calP = O(\eps^1)
\end{equation}
The fluid's force ratio is
\begin{equation}
\calQ_\ell ^f={{\frac{\mu_f V^f}{P^f \ell}}}=O\paren{\calP\calm \calV \calQ_\ell^s } = O\paren{\eps^{-1} \calm \calV}
\end{equation}
and
\begin{equation}
\calQ^f_L = \calm \calV
\end{equation}

Therefore, 
\begin{align}
\mathcal{M} V & = O(\epsilon ^{10} - \epsilon ^{1})\\
\calQ^f_\ell& = O(\epsilon ^{9} - \epsilon ^{0})
\end{align}

The choice of dimensionless parameters will lead to different
expansions and effective equations.  In the terminology of
\cite{auriault1991hme, auriault1991pm, geindreau1999ivb,
  auriault1992mhc}, we can derive one of several outcomes:
\emph{biphasic} media, \emph{monophasic} media, and
\emph{non-homogenizable} media.   In the biphasic case, the macroscopic description possesses a distinct velocity field for each phase.  In the monophasic case  \emph{both} phases have the same velocity field and we have a single, hybrid, material.  In both biphasic and monophasic models, there is only one pressure.  The non-homogenizable case is explained in Appendix \ref{app:non_homog}.

From here on, we assume
\begin{equation}
\label{eq:scaling_assumption}
\boxed{\calQ^f_\ell =O\paren{\frac{\calm \calV }{\eps}}= O(\eps)}
\end{equation}
which implies that at the microscale, the ratio of viscous stresses to
pressure in the liquid phase is $O(\eps)$.
This includes two biphasic cases, $(\calm,\calV) = (O(\eps^2), O(1))$ and $(\calm,\calV)= (O(\eps^3),O(\eps^{-1}))$, and a related monophasic case.  We discuss the significance of constraint \eqref{eq:scaling_assumption} in Section \ref{sec:homog_discuss}.

\subsection{Main Results}
\label{sec:multiscale_results}
Before proceeding with the expansions, we state our main results.  The dependent variables are $\mathbf{V}^s$, the leading order velocity in the matrix, $P$, the leading order (fluid) pressure, and $\mathbf{V}^f$, the leading order mean velocity in the fluid.  Full notation is given in Table \ref{table:effective}.

 \begin{table}
  \centering
  \caption{Effective quantities derived by homogenization.}
  \label{table:effective}
  \begin{tabular}{rp{12cm}}
  \hline\hline
  Symbol & Meaning \\
  \hline
  $\eta_\eff$ & Effective supplementary anisotropic viscosity, a fourth order tensor.  $\eta_\eff^{lm} = 2\mu_s \mean{\bar{\chi}^{lm}}_s$ is a second order tensor.\\
$\mean{K}_f$ & Permeability, a second order tensor. The i--th column is given  by $\mean{\mathbf{k}^i}_f$.\\
$k_\eff$ & Isotropic permeability.  Under symmetry, $k_\eff = \mean{\mathbf{k}^1_1}_f$\\
$\mean{\cdot}_f$ & Volume average of a quantity over the melt portion of a cell, $\mean{\cdot}_f = \int_{Y_f} \cdot d \by$\\
$\mean{\cdot}_s$ & Volume average of a quantity over the matrix portion of a cell, $\mean{\cdot}_s = \int_{Y_s} \cdot d \by$\\
$P$ & Effective macroscopic (fluid) pressure, $P = p^\fzero$.\\
$\mathbf{V}^f$ & Effective macroscopic fluid velocity, $\mathbf{V}^f = \mean{\bv^\fzero}_f / \f$.\\
$\mathbf{V}^s$ & Effective macroscopic solid velocity, $\mathbf{V}^s = \bv^{s(0)}$.\\
$\zeta_\eff$ &Effective bulk viscosity of the matrix, $\zeta_\eff =\mu_s \mean{\zeta}_s - \frac{2}{3} \mu_s (1-\f) $\\
    \hline
  \end{tabular}
\end{table}

The following systems of equations are derived in Section 
\ref{sec:expansions_details} and Appendices \ref{app:expansions}--\ref{sec:other_scalings} 
by multiple scale expansions.  They employ an additional assumption 
that the cell microstructure possesses certain symmetries which are 
discussed in Section \ref{sec:symmetry} and Appendix \ref{appendix:symmetry}.
\begin{description}
\item[Biphasic-I:]In the first biphasic case, $\calV=O(1)$ and $\calm = O(\eps^2)$, the leading order non-dimensional equations are:
\begin{subequations}
\begin{gather}
\label{eq:bi1}
\begin{split}
0&= \overline{\rho}\bg-\nabla P +\nabla\bracket{\paren{\zeta_\eff-\frac{2}{3} \mu_s(1-\f)  } \nabla \cdot \bV^s }\\
&\quad+ \nabla \cdot\bracket{2{(1-\f)\mu_s }e(\bV^s)   } + \nabla \cdot \bracket{2\eta_\eff^{lm} e_{lm}(\bV^s) }
\end{split}\\
\label{eq:bi2}
\f(\bV^f -\bV^s) = -\frac{k_\eff}{\mu_f}\paren{\nabla P - \bg^f}\\
\label{eq:bi3}
\nabla \cdot\bracket{\f \bV^f+(1-\f)\bV^s}=0
\end{gather}
\end{subequations}
Again, we emphasize that the assumption $\calV=O(1=\eps^0)$ does not
imply that the melt and solid velocities are equal, simply that the
ratio of the viscosities is of significantly different order than the
ratio of the velocities. The $lm$ terms are summed over all pairs of
$l$ and $m$.  $k_\eff$ is a scalar permeability (and more generally a
second order tensor), $\zeta_\eff$ is an effective scalar bulk
viscosity, and for each pair of indices $(l,m)$, $\eta_\eff^{lm}$ is
the anisotropic contribution to the effective shear viscosity, a
second order tensor.  These material properties, defined in terms of
microscale ``cell problems" have been simplified through the domain
symmetries. Derivatives are taken with respect to the dimensionless
macroscopic scale $\bx$, which we have suppressed as a subscript for
clarity. We note that the equations for the Biphasic-I scaling are in good
agreement with previous formulations. The chief difference is the
appearance $\eta_\eff$ term capturing the grain scale anisotropy,
which is new.

\item[Biphasic-II:]In the second biphasic case, $\calV=O(\eps^{-1})$
  and $\calm = O(\eps^3)$, the macroscopic system is:
\begin{subequations}
\begin{gather}
\begin{split}
0&= \overline{\rho}\bg-\nabla P +\nabla \bracket{\paren{\zeta_\eff-\frac{2}{3} \mu_s(1-\f)  } \nabla  \cdot \bV^s}\\
&\quad+ \nabla  \cdot\bracket{2{(1-\f)\mu_s }e(\bV^s)   } + \nabla  \cdot \bracket{2\eta_\eff^{lm} e_{lm}(\bV^s) }
\end{split}\\
\f \bV^f = -\frac{k_\eff}{\mu_f}\paren{\nabla P + \bg^f}\\
\nabla \cdot \paren{\f{\bV^f}}=0
\end{gather}
\end{subequations}
$\zeta_\eff$, $\eta_\eff$, and $k_\eff$ are as above.  The first
equation is the same as \eqref{eq:bi1} from the Biphasic-I model.  The
differences of the other two equations from (\ref{eq:bi2} --
\ref{eq:bi3}) reflect that when the fluid velocity is sufficiently
greater than the solid velocity and the viscosities are sufficiently
different,
the coupling between phases has weakened.  In this scaling regime the two phases only communicate through the pressure gradient.  

\item[Monophasic:]In the limit that the melt becomes disconnected, Biphasic-I limits to a monophasic system:
\begin{subequations}
\begin{gather}
\begin{split}
0&=\overline{\rho}\bg-\nabla P+  \nabla \cdot \bracket{2\mu_s\paren{1-\f} e(\bV^s)  +2\eta_\eff^{lm} e_{lm}(\bV^s)}
\end{split}\\
\nabla \cdot \bV^s =0
\end{gather}
\end{subequations}
where $\eta_\eff$ is as above.   This is an incompressible Stokes system modeling a composite material with an anisotropic viscosity.
\end{description}


\section{Detailed Expansions and Matching Orders}
\label{sec:expansions_details}
This section and Appendices \ref{app:expansions}--\ref{sec:matrix_cell} 
provide the detailed derivation and expansions required to 
derive the Biphasic-I model summarized in Section
\ref{sec:multiscale_results}.  The other two models are derived in Appendix \ref{sec:other_scalings}.  This material is admittedly technical but we want to provide sufficient information to offer a road map for
related studies.  

We begin by writing our equations in dimensionless form. For the scaling regimes we study, the
dimensionless forms of the force equations, \eqref{eq:scaled_force},
the incompressibility equations, \eqref{eq:inc}, and the boundary
conditions, \eqref{eq:stress_bc} and \eqref{eq:velocity_bc}, are:
\begin{subequations}
\begin{gather}
\label{eq:dimensionless_solid_force}
\nabla_y \cdot \bracket{-\tilde{p}^{s} I +2\eps^{-1} \tilde{\mu}_s e_y(\tilde{\bv}^{s} ) } +\eps\tilde{\bg}^s=0\\
\label{eq:dimensionless_fluid_force}
\nabla_y \cdot \bracket{-\tilde{p}^{f} I + 2\eps^1\tilde{\mu}_f e_y(\tilde{\bv}^{f}) }+ \eps\tilde{\bg}^f=0\\
\label{eq:dimensionless_solid_inc}
\nabla_y \cdot\tilde{\bv}^{s}=0\\
\label{eq:dimensionless_fluid_inc}
\nabla_y \cdot\tilde{\bv}^{f}=0\\
\label{eq:dimensionless_stress_bc}
\bracket{-\tilde{p}^{s} I +2\eps^{-1} \tilde{\mu}_s e_y(\tilde{\bv}^{s} ) }\cdot \bn
= \bracket{-\tilde{p}^{f} I + 2 \eps^1\tilde{\mu}_f e_y(\tilde{\bv}^{f}) }\cdot\bn\\
\label{eq:dimensionless_velocity_bc}
\tilde{\bv}^{s}= \calV\tilde{\bv}^{f}
\end{gather}
\end{subequations}

All dependent variables are functions of both $\bx$ and $\by$.  Periodicity 
in $\by$ is imposed to capture the periodicity of the microstructure.  
Derivatives act on both arguments,
\begin{equation}
\boxed{\frac{\partial}{\partial{y}_i} \mapsto \frac{\partial}{\partial{y}_i} +\eps \frac{\partial}{\partial{x}_i} }
\end{equation}
Analogously, divergence, gradient, and strain rate operators become:
\begin{subequations}
\begin{align}
\label{eq:div_twoscale}
\nabla_y\cdot &\mapsto \nabla_y\cdot + \eps \nabla_x\cdot\\
\label{eq:grad_twoscale}
\nabla_y &\mapsto \nabla_y + \eps \nabla_x\\
\label{eq:strainrate_twoscale}
e_y&\mapsto e_y + \eps e_x
\end{align}
\end{subequations}

\subsection{Hierarchy of Equations}
Expanding all variables using Eq.\ (\ref{eq:variable_expansion}) and
applying the two scale derivatives, we arrive at two hierarchies of
equations, one for each phase, which can be solved successively.
Details of these expansions are given in Appendix
\ref{app:expansions}. For the matrix, each iterate is:
\begin{subequations}
\begin{align}
\label{eq:solid_n_force}
O(\eps^n):&\quad\nabla_y \cdot \sigma^{s(n)} + \nabla_x \cdot \sigma^{s(n-1)} +\delta_{n,1}\bg^s =0\quad\text{in $Y_s$}\\
\label{eq:solid_n_inc}
O(\eps^{n+1}):&\quad\nabla_y \cdot \bv^{s(n+1)} + \nabla_x \cdot \bv^{s(n)} =0\quad\text{in $Y_s$}\\
\label{eq:solid_n_bc}
O(\eps^{n}):&\quad\sigma^{s(n)}\cdot \bn  = \sigma^{f(n)}\cdot \bn\quad\text{on $\gamma$}\\
\label{eq:solid_n_stress}
\sigma^{s(n)} &\equiv -p^{s(n)} I + 2 \mu_s\bracket{e_x(\bv^{s(n)})+e_y(\bv^{s(n+1)})}
\end{align}
\end{subequations}
where $\delta_{n,1}$ is the Kronecker delta, such that gravity only acts
at order $n=1$.  Gravity does not participate in the earlier iterates. 
Treating $\sigma^{s(n-1)}$ and $\bv^{s(n)}$ as known, the equations can be interpreted as an inhomogeneous Stokes system  for $\bv^{s(n+1)}$ and $p^{s(n)}$.  The first iterate of this system is at $n=-1$, and we set $\sigma^{s(-2)} = \bv^{s(-1)} = \sigma^{f(-1)} = p^{s(-1)}=0$.

We note that the above equations can be interpreted at each order as a
linear system in the spirit of the linear algebra problem
$\mathbf{A}\vec{x} = \vec{b}$.  As with all such problems, there is a
solvability condition which must be satisfied.  For our system, the
constraint can be interpreted as follows: to be
solvable at order $n$,  the integrated surface stress on the solid
exerted by the fluid, must match the integrated force felt within in the solid,
\begin{equation}
\label{eq:solid_solvability}
\int_\gamma \sigma^{f(n)}\cdot \bn dS = - \int_{Y_s}\paren{ \nabla_x \cdot \sigma^{s(n-1)} +\delta_{n,1}\bg^s} d\by.
\end{equation}
The enforcement of \eqref{eq:solid_solvability} separates scales and steers us to the macroscopic system.  This condition can be derived by integrating \eqref{eq:solid_n_force} over $Y_s$, invoking the divergence theorem on the $\nabla_y \cdot \sigma^{s(n)}$ term, and applying boundary condition  \eqref{eq:solid_n_bc}.


Complementing the equations for the matrix is a hierarchy of equations
for the melt phase:
\begin{subequations}
\begin{align}
\label{eq:fluid_n_force}
\begin{split}
O(\eps^{n+1}):&\quad\nabla_y \cdot \sigma^{f(n+1)}+ \nabla_x \cdot \sigma^{f(n)} +\delta_{n,1}\bg^f =0\quad\text{in $Y_f$}
 \end{split} \\
\label{eq:fluid_n_inc}
O(\eps^{n}):&\quad\nabla_y \cdot \bv^{f(n)} + \nabla_x \cdot \bv^{f(n-1)} =0\quad\text{in $Y_f$}\\
\label{eq:fluid_n_bc}
O(\eps^{n}):&\quad \bv^{f(n)} =\begin{cases} \bv^{s(n)}&\text{on $\gamma$ if $\calV = O(1)$,}\\
\bv^{s(n-1)} & \text{on $\gamma$ if $\calV = O(\eps^{-1})$.}
\end{cases}\\
\label{eq:fluid_n_stress}
\begin{split}
\sigma^{f(n)} &\equiv -p^{f(n)} I + 2 \mu_f\paren{e_y(\bv^{f(n-1)})+e_x(\bv^{f(n-2)}}
\end{split}
\end{align}
\end{subequations}
As in the solid case, we treat lower order terms, $\sigma^{f(n-1)}$ and $\bv^{f(n-2)}$ as known, then solve for pressure $p^{f(n)}$ and velocity $\bv^{f(n-1)}$.  The first iterate of this system is at $n=-1$, and we set $\sigma^{f(-1)} = \bv^{f(-1)} = \bv^{f(-2)}  = \bv^{s(-1)}=\bv^{s(-2)}=0$.

Again, there is a solvability condition.    At each order, the flow of the solid at the boundary must balance the dilation or compaction of the fluid:
\begin{equation}
\label{eq:fluid_solvability}
\int_{Y_f} \nabla_x \cdot \bv^{f(n)} d\by=\begin{cases}
- \int_\gamma\bv^{s(n)} \cdot\bn dS & \textrm{if $\calV = O(1)$}\\
- \int_\gamma  \bv^{s(n-1)} \cdot\bn dS &\textrm{if $\calV = O(\eps^{-1})$}
\end{cases} 
\end{equation}
This can be derived by integrating \eqref{eq:fluid_n_inc} over $Y_f$,
invoking the divergence theorem on the $\nabla_y \cdot \bv^{f(n)}$
term, and applying boundary condition \eqref{eq:fluid_n_bc}.
Both solvability conditions (\ref{eq:solid_solvability}) and
(\ref{eq:fluid_solvability}) will be essential for developing macroscopic
effective media equations.

\pagebreak{}
\subsection{Leading Order Equations}
\label{sec:leading_order}
The leading order equations are the same in the three scaling regimes we examine.  From \eqref{eq:solid_n_force}, \eqref{eq:solid_n_inc}, \eqref{eq:solid_n_bc}, and \eqref{eq:fluid_n_force}, the leading equations are:
\begin{subequations}
\begin{align}
\label{eq:firstorder1}
O(\eps^{-1})&:\quad \nabla_y \cdot \sigma^{s(-1)} = 0\quad\text{in $Y_s$}\\
\label{eq:firstorder3}
O(\eps ^{0})&:\quad \nabla_y \cdot \bv^\szero = 0\quad \text{in
  $Y_s$}\\
\label{eq:firstorder2}
O(\eps ^{0})&:\quad \nabla_y \cdot \sigma^\fzero = 0\quad \text{in $Y_f$}\\
\label{eq:firstorder4}
O(\eps ^{-1})&:\quad\sigma^{s(-1)}\cdot \bn = 0\quad\mbox{on $\gamma$}
\end{align}
\end{subequations}
These equations can be solved analytically to show that the leading order matrix velocity and melt pressure are independent of the fine scale.

To solve for the leading order solid velocity $\bv^\szero$, note that
the solid stress is $\sigma^{s(-1)}= 2\mu_s e_y(\bv^\szero)$ from
\eqref{eq:solid_n_stress} and multiply \eqref{eq:firstorder1} by $\bv^\szero$ and integrate by parts over $Y_s$,
\[
\begin{split}
&\int_{Y_s} \partial_{y_j} v_i^\szero\sigma_{ij}^{s(-1)}d\by\\
&\quad = \int_\gamma v_i^\szero \sigma_{ij}^{s(-1)}n_j dS - 2\mu_s\int_{Y_s} \abs{e_y(\bv^\szero)}^2d\by=0
\end{split}
\]
Applying the boundary condition \eqref{eq:firstorder4}, we obtain
\[
\int_{Y_s} \abs{e_y(\bv^\szero)}^2d\by=0.
\]
which implies that $\bv^\szero$ is constant in $\by$,
\begin{equation}
\bv ^{s(0)}= \bv^{s(0)}(\bx)
\end{equation}
$\bv^{s(0)}$ automatically satisfies \eqref{eq:firstorder3}.  

Turning to the fluid, the fluid stress is given by
\eqref{eq:fluid_n_stress} as $\sigma^\fzero = - p^\fzero I$, thus
\eqref{eq:firstorder2} becomes,
\[
\nabla_y \cdot \sigma^\fzero = \nabla_y\cdot(-p^\fzero I) = -\nabla_y p^\fzero=0.
\]
which implies,
\begin{equation}
p^{f(0)}= p^{f(0)}(\bx).
\end{equation}

\subsection{Successive Orders in the Solid Phase}
\label{sec:solidphase}

At the next order ($n=0$) in (\ref{eq:solid_n_force}--\ref{eq:solid_n_bc}),
\begin{subequations}
\begin{align}
\label{eq:solidsecondorder1}
O(\eps^{0})&:\quad \nabla_y \cdot \sigma^ \szero = 0\quad\text{in $Y_s$}\\
\label{eq:solidsecondorder2}
O(\eps ^{1})&:\quad \nabla_x \cdot \bv^ \szero +\nabla_y \cdot \bv^\sone = 0\quad\text{in $Y_s$}\\
\label{eq:solidsecondorder3}
O(\eps ^{0})&:\quad\sigma^ \szero\cdot \bn = \sigma^{f(0)}\cdot \bn\quad\text{on $\gamma$}
\end{align}
\end{subequations}
From \eqref{eq:solid_n_stress}, 
\begin{equation*}
\sigma^\szero\equiv - p^\szero I + 2 \mu_s\paren{ e_y(\bv^\sone)+e_x(\bv^\szero)}
\end{equation*}
Solvability condition \eqref{eq:solid_solvability} on the stresses is satisfied because $p^\fzero = p^\fzero(\bx)$ yielding
\[
\int_\Gamma \paren{ - p^\fzero I}\cdot \bn = 0.
\]

It is helpful to define the pressure difference between solid and
fluid as  $q = p^\szero-p^\fzero$.  $\bv^\sone$ and $q$ solve
\begin{subequations}
\begin{align}
\label{eq:solid_voneq1}
\nabla_y \cdot \paren{-q I+ 2\mu_s e_y(\bv^\sone)}&=0\quad\text{in $Y_s$}\\
\label{eq:solid_voneq2}
\nabla_y \cdot \bv^\sone &= -\nabla_x \cdot \bv^\szero\quad\text{in $Y_s$}\\
\label{eq:solid_voneq3}
\paren{-q I + 2\mu_s e_y(\bv^\sone)}\cdot \bn &= \paren{- 2 \mu_s e_x(\bv^\szero)}\cdot \bn\quad\text{on $\gamma$}
\end{align}
\end{subequations}
This is an inhomogeneous Stokes problem with the forcing terms  $\nabla_x\cdot \bv^\szero$ in \eqref{eq:solid_voneq2}  and $2 \mu_s e_x(\bv^\szero)\cdot \bn$ in \eqref{eq:solid_voneq3}; all forcing terms are independent of $\by$.  Because the problem is linear, we can solve for each component of the forcing independently.  The complete solution is the superposition:
\begin{align}
\label{eq:velocity_s1}
\begin{split}
\bv^\sone &= 2e_{x,lm}\paren{\bv^{s(0)}} \bar{\chi}^{lm} - \paren{\nabla_x \cdot \bv^{s(0)}}\bar{\xi} 
\end{split}\\
\label{eq:pressure_s0}
\begin{split}
q = p^\szero- p^\fzero&=2\mu_s e_{x,lm}\paren{\bv^{s(0)}} \pi^{lm}- \mu_s\paren{\nabla_x \cdot \bv^{s(0)}}\zeta
\end{split}
\end{align}
Summation over $lm$ is implied.  For each ordered pair $(l,m)$, there is a velocity,  $\bar{\chi}^{lm}$, and pressure,  $\pi^{lm}$, contributed from the corresponding component of the surface stress on the solid, $2 e_{x,lm}(\bv^\szero)$.  The velocity $\bar{\xi}$ and pressure $\zeta$ arise from the dilation/compaction forcing.  $\bar{\chi}^{lm}$, $\pi^{lm}$, $\bar{\xi}$, and $\zeta$ are defined in Table \ref{table:cell_problems} and full statements of the cell problems are given in Appendix \ref{sec:matrix_cell}.   These solve the aforementioned auxiliary, or cell, problems, which are Stokes boundary value problems posed on $Y_s$.  

Cell problems may be interpreted as the unit response of the medium to a particular forcing.  For generic three-dimensional cell geometries, the cell problems lack clear analytic solutions, and one must resort to numerical computation to understand them.  In our second paper, we survey them numerically.

 \begin{table}
  \centering
  \caption{Notation for cell problems.}
  \label{table:cell_problems}
  \begin{tabular}{rp{14cm}}
  \hline\hline
  Symbol & Meaning \\
  \hline
  $\bar{\chi}^{lm}$ & Velocity of the cell problem for a unit shear stress forcing on the solid in the $lm$ component of the stress tensor\\
  $\mathbf{e}_i$ & Unit vector in the i--th coordinate, $\mathbf{e}_1^T = (1, 0, 0)$\\
  $\mathbf{k}^i$& Velocity of the cell problem for a unit forcing on the fluid in the   $\mathbf{e}_i$  direction\\
$\bar{\xi}$ & Velocity of the cell problem for a unit forcing on the divergence equation\\
  $\pi^{lm}$& Pressure of the cell problem for a unit shear stress forcing on the solid in the $lm$ component of the the stress tensor\\
  $ q_i$&Pressure of the cell problem for a unit forcing on the fluid in the   $\mathbf{e}_i$ direction\\
  $\zeta$& Pressure of the cell problem for a unit forcing on the divergence equation\\
    \hline
  \end{tabular}
\end{table}

We make two observations on \eqref{eq:pressure_s0}.  First, it agrees with models that permit the pressures to be unequal, as in \cite{scott1984ms,stevenson1991mfr, bercovici2001a,  bercovici2003etp}.  It also makes clear that the question of whether there are one or two pressures in macroscopic models of partial melts is entirely semantic.  There are two pressures, but to leading order each can be expressed in terms of the other.  Second, it captures that part of any pressure jump is due to the macroscopic compaction of the matrix.  Such a relation was also discussed in \cite{spiegelman07aia, katz2007nsg}.


\subsection{Macroscopic Force Balance in the Matrix}
Though we have solved for $\bv^\sone$, $p^\szero$ in terms of $\bv^\szero$ and $p^\fzero$, we still do not have a macroscopic equation relating velocity and pressure.  To find such an equation we go to the next order of equations for the matrix and use the solvability condition, \eqref{eq:solid_solvability}, to constrain them.  This constraint becomes our macroscopic equation; we do not actually solve for $\bv^{s(2)}$ and $p^{s(1)}$.

At the next order of (\ref{eq:solid_n_force}--\ref{eq:solid_n_bc}), the equations are:
\begin{subequations}
\begin{align}
\label{eq:solidmacro1}
O(\eps^{1})&:\quad \nabla_x \cdot \sigma^{s(0)} +\nabla_y \cdot \sigma^{s(1)} + \bg^{s}=0\quad\text{in $Y_s$}\\
\label{eq:solidmacro4}
O(\eps ^{2})&:\quad \nabla_x \cdot \bv^\sone + \nabla_y\cdot \bv^{s(2)}=0 \quad\text{in $Y_s$}\\
\label{eq:solidmacro3}
O(\eps ^{1})&:\quad\sigma^{s(1)}\cdot \bn = \sigma^{f(1)}\cdot \bn\quad\text{on $\gamma$}
\end{align}
\end{subequations}
$\sigma^{s(1)}$ is given by \eqref{eq:solid_n_stress}:
\begin{equation}
\sigma^\sone =  -p^\sone + 2\mu_s \bracket{e_x(\bv^\sone)+e_y(\bv^{v(2)})}
\end{equation}

According to our force matching solvability condition \eqref{eq:solid_solvability},
\begin{equation}
\int_\gamma \sigma^\fone\cdot \bn dS = -\int_{Y_s}\paren{  \nabla_x \cdot \sigma^{s(0)}+\bg^{s}}d\by
\end{equation}
By stress boundary condition \eqref{eq:solidmacro3}, $\sigma^\sone \cdot\bn = \sigma ^\fone \cdot\bn$ on $\gamma$, so
\begin{equation}
\begin{split}
\int_{Y_s} \paren{ \nabla_x \cdot \sigma^{s(0)}+\bg^{s}}d\by&=-\int_\gamma \sigma^\fone\cdot \bn dS\\
& = \int_{Y_f} \nabla_y\cdot \sigma^\fone d\by
\end{split}
\end{equation}
Using fluid momentum equation \eqref{eq:fluid_n_force}, $\nabla_y \cdot \sigma^\fone = - \nabla_x \cdot \sigma^\fzero - \bg^f$ in $Y_f$, hence
\begin{equation}
\begin{split}
 \int_{Y_s}\paren{ \nabla_x \cdot \sigma^{s(0)} }d\by &+ \int_{Y_f}\paren{ \nabla_x \cdot \sigma^{f(0)} }d\by\\
 &\quad+ \paren{1-\f}\bg^{s} +\f\bg^{f} =0
\end{split}
\end{equation}
Commuting the integration and divergence operators,
\begin{equation}
\begin{split}
&-\nabla_x\bracket{ \mean{p^{s(0)}}_s+\mean{p^{f(0)}}_f }\\
&\quad +2\mu_s \nabla_x \cdot \bracket{\mean{e_x\paren{\bv^{s(0)}}}_s+\mean{e_y\paren{\bv^{s(1)}}}_s}\\
&\quad+\overline{\rho}\bg=0
\end{split}
\end{equation}
where angle brackets $\mean{\cdot}_s$ denote volume averages over
the solid domain $Y_s$ (likewise $\mean{\cdot}_f$ over $Y_f$).
If we substitute \eqref{eq:velocity_s1} and \eqref{eq:pressure_s0} for $p^\szero$ and $\bv^\sone$, then
\begin{equation}
\label{eq:solidmacro}
\begin{split}
0&=\overline{\rho}\bg-\nabla_x p^{f(0)}\\
&\quad - \nabla_x \set{2\mu_s e_{x,lm}\paren{\bv^{s(0)}}\mean{ \pi^{lm}}_s  -\mu_s \mean{\zeta} _s{\nabla_x \cdot \bv^{s(0)}}}\\
&\quad+ 2\mu_s \nabla_x \cdot \set{\paren{1-\f} e_x(\bv^{s(0)})  +2e_{x,lm}(\bv^{s(0)}) \mean{e_y(\bar{\chi}^{lm}) }_s}\\
&\quad -2\mu_s \nabla_x \cdot \set{\mean{e_y(\bar{\xi})}_s {\nabla_x \cdot \bv^{s(0)}}}
\end{split}
\end{equation}
We now have an equation for $\bv^\szero$ and $p^\fzero$, both functions of $\bx$.  Multiplying this equation by $P^s/L$ restores dimensions.\emph{Again, we note that we did not solve (\ref{eq:solidmacro1}--\ref{eq:solidmacro3}).}  \eqref{eq:solidmacro} is merely the equation that must be satisfied for (\ref{eq:solidmacro1}--\ref{eq:solidmacro3}) to satisfy momentum compatibility condition \eqref{eq:solid_solvability}.

\subsection{Macroscopic Force Balance in the Fluid}
\label{sec:phaseregimes}
We now seek  macroscopic equations for the melt.  As in the case of the solid, we must iterate out to the second order correction and use the solvability condition to obtain a macroscopic equation.  

We first solve for the first correction, obtaining $\bv^\fzero$ and $p^\fone$, and average them.  From the hierarchy of fluid equations, (\ref{eq:fluid_n_force} -- \ref{eq:fluid_n_stress}), the fluid equations at this order are
\begin{align}
\label{eq:fluidmacro1}
O(\eps ^{1}):&\quad \nabla_x \cdot \sigma^ \fzero +\nabla_y \cdot \sigma^\fone + \bg^{f}=0\quad\text{in $Y_f$}\\
\label{eq:fluidmacro2}
O(\eps^0): & \quad \nabla_y \cdot \bv^\fzero = 0 \quad \text{in $Y_f$}
\end{align}
with stress
\[
\sigma^\fone = -p^\fone I  + 2 \mu_f{e_y(\bv^\fzero)}
\]
and boundary conditions
\begin{equation}
\label{eq:biphasic_bc}
\bv^\fzero = \begin{cases} \bv^\szero & \text{on $\gamma$ if $\calV = O(1)$}\\
               0 \quad & \text{on $\gamma$ if $\calV = O(\eps^{-1})$}
               \end{cases}
\end{equation}

One of the most relevant scaling regimes for magma migration is Biphasic-I with $\calV=O(1)$ and $\calm=O(\eps^{2})$, summarized in Section \ref{sec:multiscale_results}.  We derive it here.  The two other systems, Biphasic-II and Monophasic, are similar and presented in Appendix \ref{sec:other_scalings}.  

If we substitute the stresses  $\sigma^\fzero$ and $\sigma^\fone$ into (\ref{eq:fluidmacro1} -- \ref{eq:fluidmacro2}), we have
\begin{align*}
-\nabla_x p^\fzero-\nabla_y p^\fone + \mu_f \nabla_y^2 \bv^\fzero + \bg^f=0\quad\text{in $Y_f$}\\
\nabla_y \cdot \bv^\fzero = 0 \quad \text{in $Y_f$}
\end{align*}
with boundary condition $\bv^\fzero = \bv^\szero$ on $\gamma$.  Recall that $p^\fzero$, $\bv^\szero$, and $\bg^f$ are interpreted as known, inhomogeneous, $\by$ independent quantities forcing $\bv^\fzero$ and $p^\fone$.  Since it is easier to solve a problem with homogeneous boundary conditions, we define $\mathbf{w} \equiv \bv^\fzero-\bv^\szero$, simplifying the above equations into
\begin{subequations}
\begin{align}
\label{eq:darcy1}
-\nabla_y p^\fone + \mu_f \nabla_y^2 \mathbf{w} = \nabla_x p^\fzero- \bg^f\quad\text{in $Y_f$}\\
\label{eq:darcy2}
\nabla_y \cdot \mathbf{w}= 0\quad \text{in $Y_f$}\\
\label{eq:darcy3}
\mathbf{w} =0\quad \text{on $\gamma$}
\end{align}
\end{subequations}
This is the classic homogenization problem of flow in a rigid porous medium and leads to Darcy's Law. It is discussed in many of the cited texts on homogenization, particularly \cite{hornung1997hap}.


The volume compatibility condition \eqref{eq:fluid_solvability} is trivially satisfied since $\mathbf{w}|_\gamma =0$,
\[
0=\int_{Y_f} (\nabla_y \cdot \mathbf{w})d\by = \int_{\gamma} \mathbf{w} \cdot \bn dS = 0.
\]

As in the case of the solid phase, we solve
\eqref{eq:darcy1}--\eqref{eq:darcy_cell3} via cell problems, taking advantage of the linearity of the problem.  We decompose the right hand side forcing terms in \eqref{eq:darcy1} into $\mathbf{e}_1$, $\mathbf{e}_2$ and $\mathbf{e}_3$ components, solving in each coordinate, then forming the superposition of the three to get the solution.  Let $q^i$, $\mathbf{k}^i$ be  $\by$ periodic functions solving:
\begin{subequations}
\begin{align}
\label{eq:darcy_cell1}-\nabla_y q^i + \nabla_y^2 \mathbf{k}^i &= -\mathbf{e}_i\quad\text{in $Y_f$}\\
\label{eq:darcy_cell2}\nabla_y \cdot \mathbf{k}^i &= 0\quad\text{in $Y_f$}\\
\label{eq:darcy_cell3}\mathbf{k}^i &= 0\quad \text{on $\gamma$}
\end{align}
\end{subequations}
$\mathbf{e}_i$ is the unit vector in the $i$-th direction.  These
problems thus measure the unit response of the fluid to such a
forcing.  Using the solutions, 
\begin{align}
\mathbf{w} &=-\frac{1}{\mu_f}\mathbf{k}^i\paren{ \partial_{x_i} p^{f(0)}-{g}_i^f}\\
p^{f(1)}(\bx,\by) &= -q^i\paren{ \partial_{x_i} p^{f(0)}-{g}_i^f} 
\end{align}
Averaging over $Y_f$, we get the macroscopic equation for the fluid,
\begin{equation}
\label{eq:biphasic1fluid}
\mean{\bv^\fzero}_f -\phi\bv^\szero = - \frac{\mean{K}_f}{\mu_f}\paren{\nabla_x p^{f(0)}- \bg^f}
\end{equation}
This is Darcy's Law with buoyancy and in a moving frame.  $\mean{K}_f$ is the permeability tensor.    $K$ is the matrix, or alternatively the second order tensor,
\begin{equation}
K = \begin{bmatrix} \mathbf{k}^1 & \mathbf{k}^2 & \mathbf{k}^3\end{bmatrix}
\end{equation}
and
\begin{equation}
\label{eq:permeability_tensor}
\mean{K}_f= \begin{bmatrix} \int_{Y_f}{\mathbf{k}^1}d\by & \int_{Y_f}{\mathbf{k}^2}d\by & \int_{Y_f}{\mathbf{k}^3}d\by\end{bmatrix}
\end{equation}
While the leading order solid velocity is $\by$-independent,  the leading order fluid velocity remains sensitive to the fine scale. For a macroscopic description, it can only be defined  as an average flux; this is the {Darcy velocity} of the fluid.  

This is not yet a closed system.  Advancing to the next order of (\ref{eq:fluid_n_force} -- \ref{eq:fluid_n_bc}), we have
\begin{align}
O(\eps ^{2}):&\quad \nabla_x \cdot \sigma^\fone +\nabla_y \cdot \sigma^{f(2)} =0\quad\text{in $Y_f$}\\
O(\eps^1): & \quad \nabla_x \cdot \bv^\fzero+ \nabla_y \cdot \bv^\fone = 0 \quad \text{in $Y_f$}\\
O(\eps^1):&\quad \bv^\fone = \bv^\sone\quad\text{on $\gamma$}
\end{align}
The solution must satisfy the volume compatibility condition \eqref{eq:fluid_solvability},
\begin{equation}
\int_{Y_f} \nabla_x \cdot \bv^\fzero = - \int_\gamma \bv^{s(0)} \cdot \bn dS=0
\end{equation}
Combining this with \eqref{eq:solidsecondorder2}, we get
\begin{equation}
\label{eq:biphasic1volume}
\nabla_x \cdot \bracket{\mean{\bv^\fzero}_f+ (1-\f) \bv^\szero}= 0
\end{equation}
This is a macroscopic volume compatibility condition.  Equations \eqref{eq:solidmacro}, \eqref{eq:biphasic1fluid}, and \eqref{eq:biphasic1volume} now form a closed system.  Dimensions may be restored to \eqref{eq:biphasic1fluid} by multiplying by $V^f$ and \eqref{eq:biphasic1volume} by $V^f/L$; a factor of $\ell^2$ will appear in front of $\mean{K}_f$, as expected.



\subsection{Symmetry Simplifications}
\label{sec:symmetry}
The macroscopic equations can be simplified  if we assume that the cell geometry is symmetric with respect to both reflections about the principal axes and rigid rotations.  Though this is a further idealization, the equations retain their essential features.

Under these two assumptions, \eqref{eq:biphasic1fluid} (for Biphasic
I) and
\eqref{eq:biphasic2fluid} (for Biphasic II) are
\begin{align}
\label{eq:permeability-symmetry1}
\mean{\bv^\fzero}_f -\phi\bv^\szero &= - \frac{k_\eff}{\mu_f}\paren{\nabla_x p^\fzero-  \bg^f}\\
\label{eq:permeability-symmetry2}
\mean{\bv^\fzero}_f &= - \frac{k_\eff}{\mu_f}\paren{\nabla_x p^{f(0)}-  \bg^f}
\end{align}
\eqref{eq:solidmacro} becomes
\begin{equation}
\label{eq:biphasic-symmetry}
\begin{split}
0&=\bar{\rho}\bg-\nabla_x p^\fzero +\nabla_x\bracket{\paren{\zeta_\eff-\frac{2}{3} \mu_s(1-\f)  } \nabla_x \cdot \bv^{s(0)}}\\
&\quad+ \nabla_x \cdot\bracket{2{(1-\f)\mu_s }e_x(\bv^{s(0)}) + 2\eta_\eff^{lm} e_{x,lm}(\bv^{s(0)})  }
\end{split}
\end{equation}
$k_\eff$, $\zeta_\eff$, and $\eta_\eff$ are defined in terms of the solutions of the cell problems:
\begin{align}
k_\eff&= \mean{K_{11}}_f\\
\zeta_\eff &= \mu_s \mean{\zeta}_s-\frac{2}{3} \mu_s(1-\f) \\
\eta_\eff^{lm}&=2 \mu_s \mean{e_y(\bar{\chi}^{lm}) }_s
\end{align}
$\eta_\eff$ is a fourth order tensor.  It is a supplementary
viscosity, capturing the grain scale anisotropy of the cell
domain. With these symmetry reductions, there are now only four
material parameters to be solved for: $\mean{K_{11}}_f$,
$\mean{\zeta}_s$, $\mean{e_{y,11}(\bar{\chi}^{11})}_s$, and
$\mean{e_{y,12}(\bar{\chi}^{12})}_s$ corresponding to the macroscopic
permeability, bulk viscosity and two effective components of an
anisotropic viscosity.  Additional details of the symmetry simplifications may be found in Appendix \ref{appendix:symmetry}.  

If we now define $\bV^s \equiv \bv^{s(0)}$, $\bV^f \equiv
\mean{\bv^\fzero}_f / \phi$, and $P \equiv p^\fzero$, and drop the $x$
subscripts from the derivatives, the above equations become
(\ref{eq:bi1} -- \ref{eq:bi3}) presented in Section \ref{sec:multiscale_results}.

%

%


\section{Discussion}
\label{sec:discussion}
We have successfully used homogenization to derive three macroscopic models for conservation of momentum
in partially  molten systems.  We now consider these models further, compare them with previous
models derived using multiphase flow methods and discuss some caveats and future directions.

\subsection{Remarks on Homogenization Models}
\label{sec:homog_discuss}
The differences amongst the three models of Section \ref{sec:multiscale_results} arise from the assumptions on two dimensionless numbers, $\calV$ and $\calm$, and the microstructure.  All three rely on the additional assumptions that \cal$Q_\ell^f = O(\eps)$ and $\calP=O(1)$.  It is helpful to write the three models as a unified set of equations:
\begin{subequations}
\begin{gather}
\label{eq:unified1}
\begin{split}
0&= \overline{\rho}\bg-\nabla P +\nabla\bracket{\paren{\zeta_\eff-\frac{2}{3} \mu_s(1-\f)  } \nabla \cdot \bV^s}\\
&\quad+ \nabla \cdot\bracket{2{(1-\f)\mu_s }e(\bV^s) + 2\eta_\eff^{lm} e_{lm}(\bV^s)   }
\end{split}\\
\label{eq:unified2}
\f(\bV^f - \calV ^{-1}\bV^s) = -\frac{k_\eff}{\mu_f}\paren{\nabla P + \bg^f}\\
\label{eq:unified3}
\nabla \cdot\bracket{\f \bV^f + \calV^{-1}(1-\f)\bV^s} =0
\end{gather}
\end{subequations}
As $\calV $ varies from $O(\eps^0)$ to $O(\eps^{-1})$, we transition
between Biphasic-I and Biphasic-II.  Letting the pore network
disconnect, $k_\eff\to 0$.  Consequently, $\bV^f  \to \calV ^{-1}
\bV^s $ in \eqref{eq:unified2}. This recovers macroscopic
incompressibility in \eqref{eq:unified3}, $\nabla \cdot \bV^s = 0$.
The divergence terms also drop from the matrix force balance equation.
Making rigorous mathematical sense of the transition between the
connected and disconnected pore network is an important open problem.
It is also interesting that the scalings do not fully describe the
macroscopic equations;  the grain scale structure can play a role. 
%
%

We return to our motivating problem, partially molten rock in the asthenosphere.  As we saw in Section \ref{sec:scalings}, for a given $\eps$, the parameters $\calV$ and $\calm$ include a range where a macroscopic description is possible.  We lose our ability to homogenize when either $\calm \calV \gg \eps^2$ or $\calm \calV \ll \eps^2$.  There may be interesting transitions here.  
That the two parameters must be related by $ \calm \calV = O(\eps^2)$ would seem a serious constraint on this approach and its applicability; however, this has  another interpretation.  

The condition on $\calm \calV $ stipulates that the length scales, viscosities, and velocities, be related by
\begin{equation}
\label{eq:h_length}
L = \ell \sqrt{\frac{\mu_s}{\mu_f}\frac{V^s}{V^f}}
\end{equation}
This also assumes $\calP=O(1)$.  This can be reinterpreted as the macroscopic length scale on which, given the viscosities and characteristic velocities of a partially molten mix we should \emph{expect} to observe a biphasic, viscously deformable, porous media.  Based on our estimates on the viscosities, velocities, and grain scale in Table \ref{table:measurements}, 
\begin{equation}
\label{eq:L_est}
L \approx 10^{-1} - 10^5  \textrm{ km}
\end{equation}

Length \eqref{eq:h_length} is similar, but not identical to the compaction length of \cite{mckenzie1984gac},
\begin{equation}
\delta_{\textrm{M84}} = \sqrt{\frac{\kappa(1-\f)(\zeta_s + \frac{4}{3} \mu_s)}{\mu_f}}
\end{equation}
The general scaling is similar as $\kappa\propto \ell^2$ therefore the
leading scaling is $\ell\sqrt{\mu_s/\mu_f}$. Nevertheless
$\delta_{\textrm{M84}}$ is porosity dependent through both
permeability, $\kappa$ and the viscosities, $\zeta_s$ and $\mu_s$,
making it dynamically and spatially varying. To understand the
variation in compaction length, it is critical to calculate both
permeabilities and viscosities that are consistent with the underlying
microstructure.  Homogenization provides this computational machinery
through the cell problems.  The companion paper calculates consistent
constitutive relations for several simple pore microstructures and suggests that in the
limit $\phi\rightarrow0$, $\delta_{\textrm{M84}}\rightarrow 0$ which
has important implications for the transition to melt-free regions.
However, $L$ is not a substitute for $\delta_{\textrm{M84}}$; such a
subsidiary length scale may also appear. 

Under the assumption that $\calV=O(1)$, \eqref{eq:h_length} also bears
resemblance to the compaction length of \cite{ricard2001b},
\begin{equation}
\delta_{\textrm{BRS01}} = \sqrt{\frac{\kappa_0 \mu_s}{\mu_f}}
\end{equation}
$\kappa_0$ is a geometric prefactor in a power law scalar permeability relationship $\kappa = \kappa_0 \f^n$ and $\kappa_0 \propto \ell^2$.

\subsection{Comparison with Existing Models}

There are several interesting and important differences between our
results and previous models derived using multiphase flow methods.
Most fundamental is that we begin with a grain scale model, assume
certain scalings, and formally derive a macroscopic model.  The
anticipated constitutive laws also emerge from these
assumptions.

In the limit of large viscosity variations, the conservation of momentum equations in
previous models can be closely identified with the Biphasic-I model,
where $\calV=O(1)$ and $\calm = O(\eps^2)$, given by equations
(\ref{eq:bi1} -- \ref{eq:bi3}), providing some validation.
Compare with \cite{mckenzie1984gac}, 
\begin{subequations}
\begin{gather}
\label{eq:mck1}
\dt \paren{\rho_f \phi} + \nabla \cdot (\rho_f \phi \bV^f )= \textrm{mass transfer}\\
\label{eq:mck2}
\dt \bracket{\rho_s (1-\phi)} + \nabla \cdot \bracket{\rho_s (1-\phi) \bV^s }= -\textrm{mass transfer}\\
\label{eq:mck3}
\f(\mathbf{V}^f - \mathbf{V}^s) = - \frac{\kappa}{\mu_f}(\nabla P - \bg^f)\\
\label{eq:mck4}
\begin{split}
0&=\meanrho \bg - \nabla P + \nabla \cdot \bracket{2(1-\f) \mu_s e(\bV^s) } + \nabla \bracket{(1-\f)(\zeta_s - \frac{2}{3}\mu_s)\nabla\cdot \bV^s}
\end{split}
\end{gather}
\end{subequations}
$\kappa$, $\mu_s$, and $\zeta_s$ are the permeability, shear viscosity, and bulk viscosity, which have unspecified dependencies on porosity.  We have reused the symbols $\mathbf{V}^f$, $\mathbf{V}^s$, and $P$, to denote the macroscopic fluid and solid velocities, and pressure.

 \begin{table}
\centering
  \caption{Additional notation for the other models.}
  \label{table:other_models}
  \begin{tabular}{rl}
  \hline\hline
  Symbol & Meaning\\
  \hline
 $C_0$& An $O(1)$ Constant\\
$\kappa$&Permeability\\
$\kappa_0$&Permeability constant for a power law permeability, $\kappa = \kappa_0 \f^n$\\
$\zeta_s$&Bulk viscosity\\
    \hline
  \end{tabular}
\end{table}

In the absence of melting and freezing, there is good agreement between the two models if we make the identifications $\zeta_\eff \equiv \zeta_s$ and $k_\eff \equiv \kappa$.  The main difference is the appearance $\eta_\eff$ term in \eqref{eq:bi1}, reflecting our consideration of a microstructure.  We emphasize that this macroscopic anisotropy is geometric in origin; the grain scale model was isotropic in each phase.  

Now we compare with \cite{bercovici2003etp}, in the ``geologically relevant limit" described by the authors in their Section 3.1.   With a bit of algebra, and using our notation, this can be written as:
\begin{subequations}
\begin{gather}
\label{eq:br1}
\dt{\phi} + \nabla \cdot (\phi \bV^f )=0\\
\label{eq:br2}
\dt{(1-\phi)} + \nabla \cdot \bracket{(1-\phi) \bV^s }= 0\\
\label{eq:br3}
\f(\mathbf{V}^f - \mathbf{V}^s) = - \frac{\kappa}{\mu_f}(\nabla P - \bg^f)\\
\label{eq:br4}
\begin{split}
0&=\meanrho \bg - \nabla P + \nabla \cdot \bracket{2(1-\f) \mu_s e(\bV^s) } + \nabla \bracket{(1-\f)(\mu_s\frac{C_0}{\f} - \frac{2}{3}\mu_s)\nabla\cdot \bV^s}\\
&\quad + \nabla \paren{\textrm{surface energy and damage}}
\end{split}
\end{gather}
\end{subequations}
In this model, $C_0$ is a dimensionless, $O(1)$ constant.  The surface
energy and damage terms, which we have not reproduced, capture
surface physics and grain deformation. In the absence of these physics, there is again good agreement
between Biphasic-I and this model if we make the identifications
$\zeta_\eff \equiv \mu_s {C_0}{\f}^{-1}$ and $k_\eff \equiv \kappa$.
As with the McKenzie model, the principal difference comes from the
$\eta_\eff$ term.  \cite{bercovici2003etp} noted that if one
eliminates mass transfer in (\ref{eq:mck1} -- \ref{eq:mck4}) and
surface physics from (\ref{eq:br1} -- \ref{eq:br4}), the two models
are identical subject to the identification $\zeta_s \equiv
\mu_s{C_0}{\f}^{-1}$.

The microscale model we homogenized,  assuming only
fluid dynamical coupling between the phases, was sufficient to generate
macroscopic equations consistent with previous models in the absence
of grain-scale surface energies.  An important open problem is to find a 
grain scale model amenable to homogenization, that includes grain scale diffusion.
One might then see a consistent macroscopic manifestation of 
these physics, which could be compared to models that 
have already attempted to included them \cite[e.g.,][]{ricard2003tpd,hiermajumder2006rgb}.

As mentioned previously, an advantage of the homogenization derivation
over the multiphase flow derivation is that there is not the same need
for closures.  In other models, one may posit and then seeks closures
for permeability, bulk viscosity, shear viscosity, and interphase
force.  These parameters might be constrained by other
information; however, this will not yield an inherently
self-consistent model.  One particularly difficult closure is the
interphase force, the force that one phase exerts on the other.  The
interphase force, which is a macroscopic re-expression of the
melt-matrix boundary conditions, is poorly constrained and non-unique.
Indeed, the model in \cite{bercovici2001a}, using one interphase force
could not replicate the model of  \cite{mckenzie1984gac}.  An equally
valid interphase force led to (\ref{eq:br1} -- \ref{eq:br4}).  As
noted, taking out the additional physics, this agrees with
(\ref{eq:mck1}--\ref{eq:mck4}).  Though this a desirable result, the
non-uniqueness of the terms remains an issue.


\subsection{Some Caveats}
\label{sec:weakness}

Homogenization provides a more rigorous method for derivation of
macroscopic equations as well as a clear mechanism for computing
critical closures.  Nevertheless, it is not foolproof and includes
its own set of assumptions whose consequences need to be understood.

For example, if the cell domains of Section \ref{sec:domains} are
independent of $\bx$, then the porosity is constant:
\[
\f = \int_{Y_f} 1 d\by.
\]
But a perfectly periodic microstructure is unrealistic.
Furthermore, once motion begins, the interface moves, likely breaking
the periodic structure.  If the domains do have $\bx$ dependence, $Y_f
= Y_f(\bx)$, then we can have $\f=\f(\bx)$.  This
introduces technical difficulties in \eqref{eq:solidmacro1}, as additional 
terms for gradients with respect to the domain should now appear.  See 
Appendix \ref{app:variations} for details.

A similar omission has been made in the poro-elastic literature
\cite[see][for a discussion]{lee1997ree,lee1997tcp1,lee1997tcp2, lee2004}.
As the elastic matrix deforms, the interface moves, changing the cell
geometry.  Earlier work \cite{auriault1991hme, hornung1997hap,
  mei1989mhp} implicitly assumed that this deformation was small
compared to the grain scale and could be ignored.  This issue also bedevils
the sintering and metallurgy papers \cite{auriault1992mhc} and
\cite{geindreau1999ivb}.  In high temperature, texturally equilibrated
systems as might be expected in the asthenosphere, grain-boundary
surface forces may help to maintain the geometry of the
micro-structure even during large deformations.  However, a consistent
homogenization would need to include these additional microscale processes.

Despite this obstacle, our equations are still of utility in several ways.  The first is that they are a macroscopic description of a constant porosity piece of material.  Such a description has not been rigorously derived before for partially molten rock. It also acts as a tool for verifying the multiphase flow models.  Taking $\f$ to be instantaneously uniform, such a model should reduce to our equations.  Under simplifications, the other models, such as \cite{mckenzie1984gac, bercovici2003etp}, are in agreement, up to the $\eta_\eff$ expression.

Another interpretation is that our models are valid when porosity
varies sufficiently slowly.  Under such an assumption, the omitted
terms  would be higher order in $\eps$ and could be justifiably
dropped.  There is a certain appeal to this;  it would not make sense
to discuss the homogenization of a material in which there were
tremendous contrasts in the porosity over short length scales.
Moreover, the typical porosity is $O(1 \%)$, so that if the porosity
parameter were also scaled, these terms may indeed be small.  Such
assumptions of slowly varying porosity underly all of the multiphase
flow derivations (and general continuum mechanics approaches).

Our final interpretation is that the equations are part of a hierarchical model for partial melts.  If we ignore melting and assume constant densities, conservation of mass can be expressed as
\begin{equation}
\dt\paren{1- \f} + \nabla \cdot \bracket{(1-\f) \bV^s}=0
\end{equation}
We might then assume that the grain matrix may be approximated by some
periodic structure at each instant.  This is consistent with
observations. 
Although the matrix deforms viscously, it retains a granular
structure.  Our equations are then treated as the macroscopic force
balances to determine $\bV^s$, and the system evolves accordingly.

Another issue with the homogenization approach is that though it 
illuminates how the effective viscosities and
the permeability arise through the cell problems,
calculating the relationship between $\zeta_\eff$, $\eta_\eff$ and
$k_\eff$  and the microstructure (e.g. porosity), requires numerically
solving the cell problems.  The companion paper,
\cite{simpson08b}, explores this, calculating
effective constitutive relationships for several idealized pore geometries.



\subsection{Open Problems}
There are several ways this work might be extended.  A natural
continuation is to model the partial melt as a  random
medium. 
This might more realistically model the pore structure of rocks. 
The equations for upscaling could also be augmented by giving the
matrix a nonlinear rheology, as in \cite{auriault1992mhc,
  geindreau1999ivb}.  This may be particularly important for magma
migration; a nonlinear matrix rheology was needed to computationally
model physical experiments for shear bands in \cite{katz2006dma} and
in general, non-linear power-law rheologies are expected in the
dislocation creep regime \cite[e.g.,][]{hirth1995ecd2}.



Important mechanisms absent from these equations are
surface physics which. In a fluid dynamical description, these might take the
form as surface tension and diffusional terms.  Such terms were
posited in the models
of \cite{ricard2001b,hiermajumder2006rgb,bercovici2005tpg,bercovici2003etp,bercovici2001c,bercovici2001a},
but it remains to be shown how such terms in the macroscopic equations
might arise consistently from microscopic physics that includes
grain-scale diffusion and/or mass transfer.

The most serious question remains how to a properly study a 
medium with macroscopic and time dependent variations in the 
structure.  This would have implications for the many physical 
phenomena that also have \emph{evolving microstructures}.  
Recent work in \cite{peter2007hcd,peter2007hde,peter2008crd} on 
reaction-diffusion systems in porous media may be applicable.


%


\appendix

\section{Details of the Expansions}
\label{app:expansions}
The multiple scale expansions of \eqref{eq:variable_expansion} are
applied to $\tilde{p}^{s}$, $\tilde{p}^{f}$, $\tilde{\bv}^{s}$,
$\tilde{\bv}^f$ and substituted into the dimensionless equations (\ref{eq:dimensionless_solid_force} -- \ref{eq:dimensionless_velocity_bc}), along with the two scale derivatives, (\ref{eq:div_twoscale} -- \ref{eq:strainrate_twoscale}).  Dropping tildes, both melt and matrix strain rate tensors expand as:
\begin{equation}
\label{eq:strain_expansion}
\begin{split}
e_y(\bv) &\mapsto \eps^{0} \bracket{e_y(\bv^{(0)})} + \eps^1 \bracket{ e_x(\bv^{(0)})+ e_y(\bv^{(1)}) } + \eps ^2 \bracket{ e_x(\bv^{(1)})+ e_y(\bv^{(2)}) }+\ldots\\
&\quad\equiv \eps^{0}e^{(0)} + \eps^1 e^{(1)}+ \eps ^2 e^{(2)}+\ldots
\end{split}
\end{equation}
The stress tensors become:
\begin{subequations}
\begin{gather}
\label{eq:solid_stress_expansion}
\begin{split}
\sigma^{s} & \mapsto \eps ^{-1} \bracket{ 2 \mu_s e^{s(0)}} + \eps^0\bracket{-p^\szero I + 2 \mu_s e^\sone }+ \eps^1 \bracket{-p^\sone I + 2 \mu_s e^{s(2)} }+\ldots\\
&\quad\equiv\eps^{-1}\sigma^{s(-1)} + \eps ^{0}\sigma^\szero+ \eps ^{1}\sigma^\sone+\ldots
\end{split}\\
\label{eq:fluid_stress_expansion}
\begin{split}
\sigma^{f} & \mapsto \eps^0 \bracket{-p^\fzero I}+\eps^1 \bracket{-p^\fone I+ 2\mu_f e^\fzero }+ \eps^2\bracket{ -p^{f(2)}I+2\mu_f e^\fone}+\ldots \\
&\quad \equiv\sigma^\fzero + \eps ^{1}\sigma^\fone+ \eps ^{2}\sigma^{f(2)}+\ldots
\end{split}
\end{gather}
\end{subequations}

Matching powers of $\eps$ in equations  \eqref{eq:solid_stress_expansion} and \eqref{eq:fluid_stress_expansion} in \eqref{eq:dimensionless_solid_force} and \eqref{eq:dimensionless_fluid_force}
\begin{subequations}
\begin{gather}
\label{eq:solid-force-matched}
\begin{split}
\eps^{-1}\nabla_y \cdot\sigma^{s(-1)} &+ \eps^{0}\paren{\nabla_x \cdot \sigma^{s(-1)} + \nabla_y \cdot \sigma^{s(0)}} + \eps^{1}\paren{\nabla_x \cdot \sigma^{s(0)} + \nabla_y\cdot \sigma^{s(1)}+\bg^s}\\
&\quad+\ldots  = 0
\end{split}\\
\label{eq:fluid-force-matched}
\begin{split}
\eps^{0}\nabla_y \cdot\sigma^{f(0)} &+ \eps^{1}\paren{\nabla_x \cdot \sigma^{f(0)} + \nabla_y \cdot \sigma^{f(1)} + \bg^f} + \eps^{2}\paren{\nabla_x \cdot \sigma^{f(1)} + \nabla_y\cdot \sigma^{f(2)}}\\
& \quad +\ldots  = 0
\end{split}
\end{gather}
\end{subequations}

Analogously, we substitute the expansions into the incompressibility equations (\ref{eq:dimensionless_solid_inc}--\ref{eq:dimensionless_fluid_inc}), to get
\begin{equation}
\label{eq:inc_matched}
\begin{split}
\eps^0\nabla_y \cdot \bv^{f(0)} &+ \eps^1\paren{\nabla_x \cdot \bv^{f(0)} + \nabla_y \cdot \bv^{f(1)}}+\ldots =0\\
\eps^0\nabla_y \cdot \bv^{s(0)} &+ \eps^1\paren{\nabla_x \cdot \bv^{s(0)} + \nabla_y \cdot \bv^{s(1)}}+\ldots =0
\end{split}
\end{equation}
The leading order equations of \eqref{eq:inc_matched}, $\nabla_y \cdot \bv^{f(0)} = 0$ and $\nabla_y \cdot \bv^{s(0)} = 0$, reflect that at the grain scale, both phases are incompressible.

Making the same power series expansions in the boundary conditions, continuity of normal stress, \eqref{eq:dimensionless_stress_bc}, is
\begin{equation}
\label{eq:stress_matched}
\eps^{-1}\sigma^{s(-1)}\cdot \bn + \eps^{0} \paren{\sigma^\szero-\sigma^\fzero}\cdot \bn +\ldots = 0.
\end{equation}
When $\calV=O(1)$, the velocity boundary condition within the cell is
\begin{equation}
\label{eq:velocity_matched}
\eps^0( \bv^\szero - \bv^\fzero) + \eps^1( \bv^\sone - \bv^\fone)+\ldots =0\quad\text{on $\gamma$}.
\end{equation}
In this case, the velocities are matched at all orders of $\eps$.  If instead $\calV=O(\eps^{-1})$, then
\begin{equation}
\label{eq:velocity_unmatched}
\begin{split}
\eps^{-1}\bv^\fzero &+ \eps^{0}\paren{\bv^\fone-\bv^\szero}\\
& + \eps^{1} \paren{\bv^{f(2)}-\bv^\sone}+\ldots=0\quad\text{on $\gamma$}.
\end{split}
\end{equation}
In contrast to the $\calV=O(1)$ case, the leading order fluid velocity is independent of the solid, and there is cross coupling across orders.

\section{Cell Problems in the Matrix}
\label{sec:matrix_cell}
In general, there are two classes of cell problems associated with the matrix phase, and a total of seven cell problems.  Domain symmetry can reduce the number of unique cell problems.


\subsection{Cell Problem for Dilation Stress on Solid}
\label{sec:dilation_cell}
This addresses the term $\nabla_x \cdot \bv^{s(0)}$ in  \eqref{eq:solid_voneq2}.  This is a less common Stokes problem, with a prescribed function in the divergence equation.  They are briefly discussed in \cite{temam2001nse}.  Let  $\bar{\xi}$, $\zeta$ be $\by$ periodic functions solving 
\begin{subequations}
\begin{align}
\label{eq:dilation2}
\nabla_{y}\cdot\paren{-\zeta I + 2 e_{y}(\bar{\xi})}&=0\quad\text{in $Y_s$}\\
\label{eq:dilation3}
\nabla_y \cdot \bar{\xi}&= 1\quad \text{in $Y_s$}\\
\label{eq:dilation4}
\paren{-\zeta I + 2 e_{y}(\bar{\xi})}\cdot \bn& =0\quad \text{on $\gamma$}
\end{align}
\end{subequations}
The solution measures the response of a unit cell of the matrix to the divergence condition \eqref{eq:dilation3}.

\subsection{Cell Problem for Surface Stresses on Solid}
\label{sec:tensor_cell}
This problem tackles the boundary stress in \eqref{eq:solid_voneq3}.  Let $\bar{\chi}^{lm}$, $\pi^{lm}$ be $\by$ periodic functions solving
\begin{subequations}
\begin{align}
\label{eq:surface_tensor2}
\nabla_y\cdot\paren{-\pi^{lm}I + 2 e_{y}(\bar{\chi}^{lm})}&=0\quad\text{in $Y_s$}\\
\label{eq:surface_tensor3}
\nabla_y \cdot \bar{\chi}^{lm}&= 0\quad \text{in $Y_s$}\\
\label{eq:surface_tensor4}
\paren{-\pi^{lm}\delta_{ij} + 2 e_{y,ij}(\bar{\chi}^{lm})}n_j &=-\frac{1}{2}\paren{\delta_{il}\delta_{jm}+\delta_{im}\delta_{jl}} n_j\quad \text{on $\gamma$}
\end{align}
\end{subequations}
$(\bar{\chi}^{lm}, \pi^{lm})$ measure the response of a unit cell of
the matrix to a given unit surface stress, depending on indices $(l,m)$.  Observe that because the tensor on the right hand side of \eqref{eq:surface_tensor4}, operating on $\bn$, is symmetric, the solution to problem $(l,m)$ is the same as the solution for problem $(m,l)$.  For general domains, there are thus six unique cell problems associated with surface stress forcing.

\section{Additional Scaling Regimes}
\label{sec:other_scalings}
In addition to the Biphasic-I regime which we derived in Section 
\ref{sec:phaseregimes}, we presented two other cases in Section 
\ref{sec:multiscale_results}.  These are Biphasic-II, where  $\calV = O(\eps^{-1})$ 
and $\mathcal{M} = O(\eps^{3})$, and Monophasic, where assumes 
$\calV=O(1)$ and $\calm=O(\eps^{2})$ and additionally assumes 
the melt network is disconnected.  Their derivation is given in the next two sections.

\subsection{Biphasic-II: Unequal Velocities at the Interface}
\label{sec:biphasic2}
In this case $\calV = O(\eps^{-1})$ and $\mathcal{M} = O(\eps^{3})$. Following the scheme of Section \ref{sec:phaseregimes}, the macroscopic equations are
\begin{align}
\label{eq:biphasic2fluid}
\mean{\bv^\fzero}_f &= - \frac{\mean{K}_f}{\mu_f} \paren{\nabla_x p^\fzero- \bg^f}\\
\label{eq:biphasic2volume}
\nabla_x \cdot\mean{\bv^\fzero}_f &= 0.
\end{align}
Multiplying \eqref{eq:biphasic2fluid} by $V^f$ and \eqref{eq:biphasic2volume} by $V^f/L$ restores the dimensions of these equations.  

\subsection{Monophasic: Magma Bubbles}
\label{sec:monophasic_inc}
As in Biphasic-I, we take  $\calV=O(1)$ and $\calm=O(\eps^{2})$.  However, we now assume that the fluid is not topologically connected.The equations are the same at all orders of $\eps$ as those appearing in Section \ref{sec:phaseregimes}.  

Under this assumption on the microscopic geometry,  the permeability cell problems, (\ref{eq:darcy_cell1}--\ref{eq:darcy_cell3}), can be shown to have trivial solutions.  $\mathbf{k}^i=0$ for $i=1,2,3$, so $\mean{k}_f = 0$.  Because the melt is trapped it must migrate with the matrix,
\begin{equation}
\label{eq:trapped2}
{\bv}^{\fzero}(\bx,\by) = \bv^{\szero}(\bx)
\end{equation}
Combining \eqref{eq:trapped2} with \eqref{eq:biphasic1volume}, recovers the incompressibility of the matrix,
\begin{equation}
\label{eq:mono_inc}
\nabla_x \cdot \bv^\szero = 0
\end{equation}
Dropping the divergence terms from \eqref{eq:solidmacro} completes the system:
\begin{equation}
\label{eq:mono_force}
\begin{split}
0&=\overline{\rho}\bg-\nabla_x p^\fzero  - \nabla_x \bracket{2\mu_s e_{x,lm}(\bv^{s(0)})\mean{ \pi^{lm}}_s}\\
&\quad+ 2\mu_s \nabla_x \cdot \bracket{\paren{1-\f} e_x(\bv^{s(0)})  +2e_{x,lm}(\bv^{s(0)}) \mean{e_y(\bar{\chi}^{lm}) }_s}
\end{split}
\end{equation}
This is a homogenized incompressible Stokes system for a hybrid material with isolated very low viscosity inclusions. 


\section{Non-Homogenizable Regimes}
\label{app:non_homog}
When either $Q^f_\ell \gg \eps$ or $Q^f_\ell \ll \eps$, the system is non-homogenizable.  By this we mean that it is not possible to upscale equations that faithfully preserve our physical assumptions.   For instance, if $Q^f_\ell = O(1)$ the pressure gradient balances the viscous forces in the fluid and there is no scale separation.  Working out the expansions, the leading order velocity and pressure in the fluid solve:
\begin{gather}
\nabla_y \cdot \bracket{ - \tilde{p}^\fzero + 2 \tilde\mu_f e_y(\tilde\bv^\fzero)}=0\quad \text{in $Y_f$}\\
\nabla_y \cdot \tilde\bv^\fzero = 0,\quad\text{in $Y_f$}\\
\tilde\bv^\fzero = 0,\quad\text{on $\gamma$}
\end{gather}
The solution is $\tilde\bv^\fzero=0$.  Therefore,
\begin{equation}
\begin{split}
\bv^{f,\eps}&= V^f \tilde\bv^{f,\eps}\\
& = V^f \paren{\tilde\bv^\fzero + \eps \tilde\bv^\fone + \ldots}\\
&= \eps V^f\paren{\bv^\fone+\ldots}
\end{split}
\end{equation}
This implies that $\abs{\bv^{f,\eps}} =O(\eps V^f)$, contradicting our physical assumption that $\abs{\bv^{f,\eps}} = O(V^f)$.   While this is mathematically reasonable, the model is unable to produce macroscopic fluid velocities of order $V^f$.  Other upscaling techniques may succeed here, but homogenization will not.

Suppose instead $Q^f_\ell = O(\eps^2)$ or smaller. The fluid equations are then:
\begin{align}
O(\eps^0):&\quad-\nabla_y \tilde p^{f(0)} =0 \quad \text{in $Y_f$}\\
O(\eps^1):&\quad-\nabla_y \tilde p^{f(1)} -\nabla_x \tilde p^{f(0)} + \tilde\bg^f = 0\quad\text{in $Y_f$}
\end{align}
The first equation implies $\tilde p^\fzero= \tilde p^\fzero(\bx)$.  Since $\nabla_x \tilde p^\fzero$ and $\tilde\bg^f$ are independent of $\by$, $\nabla_y \tilde p^\fone$ must also be independent.  Since it is periodic in $\by$, it is zero.  But this implies
\begin{equation}
\label{eq:hydrostatic_balance}
-\nabla_x \tilde p^{f(0)} + \tilde \bg^f = 0
\end{equation}
\emph{The leading order macroscopic pressure gradient plays no role in balancing the viscous forces in the solid.}  This contradicts our assumption that there is \emph{always} a leading order non-hydrostatic pressure gradient.

Though our assumption on the non-hydrostatic pressure gradient may seem arbitrary, there is another important reason to identify cases without such a pressure as non-homogenizable.  There are problems of interest where gravity plays little role, such as \cite{spiegelman2003lam,katz2006dma}.  In these cases, $\tilde \bg^f$ would be absent from our equations, including \eqref{eq:hydrostatic_balance}.  Hence,
\begin{equation*}
\begin{split}
\nabla p^{f} &= \frac{P^f}{L}\nabla_x  \paren{\tilde p^\fzero + \eps \tilde p^\fone +\ldots}\\
&=\eps \frac{P^f}{L} \nabla_x\paren{\tilde p^\fone +\ldots}\\
& = O(\eps \frac{P^f}{L})
\end{split}
\end{equation*}
This implies the macroscopic fluid pressure gradient is not $O(P^f/L)$, as hypothesized.

\section{Cell Problem Symmetries}
\label{appendix:symmetry}
Let us assume our cell domain is symmetric with respect to the principal axes and invariant under rigid rotations.  This permits simplifications of some of the cell problems.

In the Darcy cell problem, the off-diagonal entries become zero while the diagonal entries are all equal.  Thus:
\begin{equation}
\label{eq:effective-perm}
k_\eff= \mean{\bk_1^1}_f
\end{equation}

For the surface stress problems, when $l \neq m$, $\mean{\pi^{lm}}_s=0$.  Only the $l,m$ and $m,l$ entries of the tensor $\mean{e_y(\bar{\chi} ^{lm})}_s$ are non-zero.   For $l = m$, $\mean{\pi^{ll}}_s = \frac{1}{3}(1-\f)$ and only the diagonal entries of $\mean{e_y(\bar{\chi}^{ll})}_s$ are non-zero.  The trace of all $\mean{e_y(\bar{\chi}^{lm})}_s$ tensors is zero.  More can be said about $e_{y}(\bar{\chi}^{lm})$, but it does not benefit the present analysis.  See \cite{simpson08t} or \cite{simpson08b} for more details.

In the dilation stress problem, the off diagonal terms in $\mean{e_y(\bar{\xi})}_s$ vanish, and the diagonal entries are equal to $\frac{1}{3}(1-\f)$. 

\section{Spatial Variation in Cell Domain and Time Dynamics}
\label{app:variations}
If the cells have $\bx$ dependence, $Y_f = Y_f(\bx)$, then it is possible that $\f=\f(\bx)$.  This introduces difficulties in \eqref{eq:solidmacro1}, as terms from gradients with respect to the domain now appear.  Let us elaborate.  For fixed $\bx\in \Omega $, we associate a particular cell $Y=Y(\bx)$, with fluid and solid regions defined by the indicator functions $\mathbb{I}_f $ and $\mathbb{I}_s $:
\begin{subequations}
\begin{align}
\mathbb{I}_s & : \Omega\times Y \mapsto \set{0,1}\\
\mathbb{I}_f & : \Omega\times Y \mapsto \set{0,1}
\end{align}
\end{subequations}
Then
\begin{subequations}
\begin{align}
Y_f(\bx)&= \set{\by \in Y \mid \mathbb{I}_f(\bx,\by)=1}\\
Y_s(\bx)&= \set{\by \in Y \mid \mathbb{I}_s(\bx,\by)=1}
\end{align}
\end{subequations}

Returning to \eqref{eq:solidmacro1},
\begin{equation}
\begin{split}
&\int_{Y_s} \nabla_x \cdot \sigma^\szero d \by +\int_{Y_f} \nabla_x \cdot \sigma^\fzero d \by\\
&\quad= \int_{Y} \nabla_x \cdot \sigma^\szero \mathbb{I}_s  d \by +\int_{Y} \nabla_x \cdot \sigma^\fzero \mathbb{I}_f  d \by\\
& \quad =  \nabla_x \cdot  \int_{Y_s} \sigma^\szero d \by - \int_{Y} \sigma^\szero\cdot \nabla_x \mathbb{I}_s d \by\\
&\quad \quad +\nabla_x \cdot  \int_{Y_f} \sigma^\fzero d \by - \int_{Y} \sigma^\fzero\cdot \nabla_x \mathbb{I}_f d \by
\end{split}
\end{equation}
Witness the appearance of the $\nabla\mathbb{I}_s$ and $\nabla\mathbb{I}_f$ terms.  \emph{This is only an issue for \eqref{eq:solidmacro1}.  The other macroscopic equations remain valid when we allow cell variation.}


A second problem is manifest when we consider time dynamics.
\[
\begin{split}
\dt \f  &= \dt \int_{Y_f} 1 d \by = \int_\Gamma \bv^{f}\cdot \bn dS\\
&= - \int_\Gamma \bv^{s}\cdot \bn dS = -  \int_\Gamma \paren{\bv^\szero+\eps\bv^\sone + \ldots}\cdot \bn dS
\end{split}
\]
Since $\bv^\szero$ is independent of $\by$, the first term drops.  Substituting \eqref{eq:velocity_s1},
\[
\dt \f = -\eps \int \nabla_y\cdot \bv^\sone d\by + O(\eps^2) = \eps \nabla_x\cdot \bv^\szero(1-\f)+O(\eps^2)
\]
To leading order, the matrix can only vary by dilation and compaction.

%
%
%
%
%
%

%
%
%
%

\begin{acknowledgments}
Both this paper and \cite{simpson08b} are based on the thesis of G.~Simpson, \cite{simpson08t}, completed in partial fulfillment of the requirements for the degree of doctor of philosophy at Columbia University.

The authors wish to thank  D.~Bercovici and R.~Kohn for their helpful comments.

This work was funded in part by the US National Science Foundation (NSF) Collaboration in Mathematical Geosciences (CMG), Division of Mathematical Sciences (DMS), Grant DMS--05--30853, the NSF Integrative Graduate Education and Research Traineeship (IGERT) Grant DGE--02--21041, NSF Grants DMS--04--12305 and DMS--07--07850.
\end{acknowledgments}

%
%

\bibliography{homogenization_agu_pt1_rev}

\begin{thebibliography}{55}
\providecommand{\natexlab}[1]{#1}
\expandafter\ifx\csname urlstyle\endcsname\relax
  \providecommand{\doi}[1]{doi:\discretionary{}{}{}#1}\else
  \providecommand{\doi}{doi:\discretionary{}{}{}\begingroup
  \urlstyle{rm}\Url}\fi

\bibitem[{\textit{Auriault}(1987)}]{auriault1987ndp}
Auriault, J., {Nonsaturated deformable porous media: Quasistatics},
  \textit{Transport in Porous Media}, \textit{2}(1), 45--64, 1987.

\bibitem[{\textit{Auriault}(1991{\natexlab{a}})}]{auriault1991hme}
Auriault, J., {Heterogeneous medium. Is an equivalent macroscopic description
  possible?}, \textit{International Journal of Engineering Science},
  \textit{29}(7), 785--795, 1991{\natexlab{a}}.

\bibitem[{\textit{Auriault}(1991{\natexlab{b}})}]{auriault1991pm}
Auriault, J., Poroelastic media, in \textit{{Homogenization and Porous Media}},
  edited by U.~Hornung, pp. 163--182, Springer, 1991{\natexlab{b}}.

\bibitem[{\textit{Auriault and Boutin}(1992)}]{auriault1992dpm}
Auriault, J., and C.~Boutin, {Deformable porous media with double porosity.
  Quasi-statics. I: Coupling effects}, \textit{Transport in Porous Media},
  \textit{7}(1), 63--82, 1992.

\bibitem[{\textit{Auriault and Royer}(2002)}]{auriault2002swf}
Auriault, J., and P.~Royer, {Seismic waves in fractured porous media},
  \textit{Geophysics}, \textit{67}, 259, 2002.

\bibitem[{\textit{Auriault et~al.}(1992)\textit{Auriault, Bouvard, Dellis, and
  Lafer}}]{auriault1992mhc}
Auriault, J., D.~Bouvard, C.~Dellis, and M.~Lafer, {Modelling of Hot Compaction
  of Metal Powder by Homogenization}, \textit{Mechanics of Materials(The
  Netherlands)}, \textit{13}(3), 247--255, 1992.

\bibitem[{\textit{Bensoussan et~al.}(1978)\textit{Bensoussan, Lions, and
  Papanicolaou}}]{bensoussan1978aap}
Bensoussan, A., J.~Lions, and G.~Papanicolaou, \textit{{Asymptotic Analysis for
  Periodic Structures}}, \textit{Studies in Mathematics and its Applications},
  vol.~5, Elsevier, 1978.

\bibitem[{\textit{Bercovici and Ricard}(2003)}]{bercovici2003etp}
Bercovici, D., and Y.~Ricard, {Energetics of a two-phase model of lithospheric
  damage, shear localization and plate-boundary formation}, \textit{Geophysical
  Journal International}, \textit{152}(3), 581--596, 2003.

\bibitem[{\textit{Bercovici and Ricard}(2005)}]{bercovici2005tpg}
Bercovici, D., and Y.~Ricard, {Tectonic plate generation and two-phase damage:
  Void growth versus grain size reduction}, \textit{Journal of Geophysical
  Research}, \textit{110}(B3), 2005.

\bibitem[{\textit{Bercovici et~al.}(2001{\natexlab{a}})\textit{Bercovici,
  Ricard, and Schubert}}]{bercovici2001a}
Bercovici, D., Y.~Ricard, and G.~Schubert, {A two-phase model for compaction
  and damage, 1: General theory}, \textit{Journal of Geophysical Research},
  \textit{106}(B5), 8887--8906, 2001{\natexlab{a}}.

\bibitem[{\textit{Bercovici et~al.}(2001{\natexlab{b}})\textit{Bercovici,
  Ricard, and Schubert}}]{bercovici2001c}
Bercovici, D., Y.~Ricard, and G.~Schubert, {A Two-Phase Model for Compaction
  and Damage, 3: Applications to Shear Localization and Plate Boundary
  Formation}, \textit{Journal of Geophysical Research}, \textit{106},
  8925--8939, 2001{\natexlab{b}}.

\bibitem[{\textit{Brennen}(2005)}]{brennen2005fmf}
Brennen, C., \textit{{Fundamentals of Multiphase Flow}}, Cambridge University
  Press, 2005.

\bibitem[{\textit{Chechkin et~al.}(2007)\textit{Chechkin, Piatnitski, and
  Shamaev}}]{checkin2007hma}
Chechkin, G., A.~Piatnitski, and A.~Shamaev, \textit{{Homogenization: Methods
  and Applications}}, {Translations of Mathematical Monographs}, American
  Mathematical Society, 2007.

\bibitem[{\textit{Cioranescu and Donato}(1999)}]{cioranescu1999ih}
Cioranescu, D., and P.~Donato, \textit{{An Introduction to Homogenization}},
  Oxford University Press, 1999.

\bibitem[{\textit{Drew}(1971)}]{drew1971afe}
Drew, D., {Averaged field equations for two-phase media}, \textit{Stud. Appl.
  Math}, \textit{50}(2), 133--166, 1971.

\bibitem[{\textit{Drew}(1983)}]{drew1983mmp}
Drew, D., {Mathematical-modeling of 2-phase flow}, \textit{Ann. Rev. Fluid
  Mech}, \textit{15}, 261--291, 1983.

\bibitem[{\textit{Drew and Passman}(1999)}]{drew1999tmf}
Drew, D., and S.~Passman, \textit{{Theory of multicomponent fluids}}, Springer
  New York, 1999.

\bibitem[{\textit{Drew and Segel}(1971)}]{drew1971aet}
Drew, D., and L.~Segel, {Averaged equations for two-phase flows}, \textit{Stud.
  Appl. Math}, \textit{50}(3), 205--231, 1971.

\bibitem[{\textit{Fowler}(1985)}]{fowler1985mmm}
Fowler, A., {A mathematical model of magma transport in the asthenosphere},
  \textit{Geophysical \& Astrophysical Fluid Dynamics}, \textit{33}(1), 63--96,
  1985.

\bibitem[{\textit{Fowler}(1989)}]{fowler1989gac}
Fowler, A., {Generation and Creep of Magma in the Earth}, \textit{SIAM Journal
  on Applied Mathematics}, \textit{49}, 231, 1989.

\bibitem[{\textit{Geindreau and Auriault}(1999)}]{geindreau1999ivb}
Geindreau, C., and J.~Auriault, {Investigation of the viscoplastic behaviour of
  alloys in the semi-solid state by homogenization}, \textit{Mechanics of
  Materials}, \textit{31}(8), 535--551, 1999.

\bibitem[{\textit{Hier-Majumder et~al.}(2006)\textit{Hier-Majumder, Ricard, and
  Bercovici}}]{hiermajumder2006rgb}
Hier-Majumder, S., Y.~Ricard, and D.~Bercovici, {Role of grain boundaries in
  magma migration and storage}, \textit{Earth and Planetary Science Letters},
  \textit{248}(3-4), 735--749, 2006.

\bibitem[{\textit{Hirth and Kohlstedt}(1995{\natexlab{a}})}]{hirth1995ecd1}
Hirth, G., and D.~Kohlstedt, {Experimental constraints on the dynamics of the
  partially molten upper mantle: Deformation in the diffusion creep regime},
  \textit{Journal of Geophysical Research}, \textit{100}(B2), 1981--2001,
  1995{\natexlab{a}}.

\bibitem[{\textit{Hirth and Kohlstedt}(1995{\natexlab{b}})}]{hirth1995ecd2}
Hirth, G., and D.~Kohlstedt, {Experimental constraints on the dynamics of the
  partially molten upper mantle 2: Deformation in the dislocation creep
  regime}, \textit{Journal of Geophysical Research}, \textit{100}(B2),
  15,441--15,449, 1995{\natexlab{b}}.

\bibitem[{\textit{Hornung}(1997)}]{hornung1997hap}
Hornung, U., \textit{{Homogenization and Porous Media}}, Springer, 1997.

\bibitem[{\textit{Katz et~al.}(2006)\textit{Katz, Spiegelman, and
  Holtzman}}]{katz2006dma}
Katz, R., M.~Spiegelman, and B.~Holtzman, {The dynamics of melt and shear
  localization in partially molten aggregates.}, \textit{Nature},
  \textit{442}(7103), 676--9, 2006.

\bibitem[{\textit{Katz et~al.}(2007)\textit{Katz, Knepley, Smith, Spiegelman,
  and Coon}}]{katz2007nsg}
Katz, R., M.~Knepley, B.~Smith, M.~Spiegelman, and E.~Coon, {Numerical
  simulation of geodynamic processes with the Portable Extensible Toolkit for
  Scientific Computation}, \textit{Physics of the Earth and Planetary
  Interiors}, \textit{163}(1-4), 52--68, 2007.

\bibitem[{\textit{Lee}(2004)}]{lee2004}
Lee, C., {Flow and deformation in poroelastic media with moderate load and weak
  inertia}, \textit{Proceedings of the Royal Society of London. Series A,
  Mathematical and Physical Sciences}, \textit{460}(2047), 2051--2087, 2004.

\bibitem[{\textit{Lee and Mei}(1997{\natexlab{a}})}]{lee1997ree}
Lee, C., and C.~Mei, {Re-examination of the equations of poroelasticity},
  \textit{International journal of engineering science}, \textit{35}(4),
  329--352, 1997{\natexlab{a}}.

\bibitem[{\textit{Lee and Mei}(1997{\natexlab{b}})}]{lee1997tcp1}
Lee, C., and C.~Mei, {Thermal consolidation in porous media by homogenization
  theory---I. Derivation of macroscale equations}, \textit{Advances in Water
  Resources}, \textit{20}(2-3), 127--144, 1997{\natexlab{b}}.

\bibitem[{\textit{Lee and Mei}(1997{\natexlab{c}})}]{lee1997tcp2}
Lee, C., and C.~Mei, {Thermal consolidation in porous media by homogenization
  theory---II. Calculation of effective coefficients}, \textit{Advances in
  Water Resources}, \textit{20}(2-3), 145--156, 1997{\natexlab{c}}.

\bibitem[{\textit{McKenzie}(1984)}]{mckenzie1984gac}
McKenzie, D., The generation and compaction of partially molten rock,
  \textit{Journal of Petrology}, \textit{25}(3), 713--765, 1984.

\bibitem[{\textit{Mei and Auriault}(1989)}]{mei1989mhp}
Mei, C., and J.~Auriault, {Mechanics of Heterogeneous Porous Media With Several
  Spatial Scales}, \textit{Proceedings of the Royal Society of London. Series
  A, Mathematical and Physical Sciences}, \textit{426}(1871), 391--423, 1989.

\bibitem[{\textit{Mei et~al.}(1996)\textit{Mei, Auriault, and Ng}}]{mei1996sah}
Mei, C., J.~Auriault, and C.~Ng, {Some applications of the homogenization
  theory}, \textit{Advances in applied mechanics}, \textit{32}, 277--348, 1996.

\bibitem[{\textit{Pavliotis and Stuart}(2008)}]{pavliotis2008mma}
Pavliotis, G., and A.~Stuart, \textit{{Multiscale Methods: Averaging and
  Homogenization}}, Springer, 2008.

\bibitem[{\textit{Peter}(2007{\natexlab{a}})}]{peter2007hcd}
Peter, M., {Homogenisation of a chemical degradation mechanism inducing an
  evolving microstructure}, \textit{Comptes rendus-M{\'e}canique},
  \textit{335}(11), 679--684, 2007{\natexlab{a}}.

\bibitem[{\textit{Peter}(2007{\natexlab{b}})}]{peter2007hde}
Peter, M., {Homogenisation in domains with evolving microstructure},
  \textit{Comptes rendus-M{\'e}canique}, \textit{335}(7), 357--362,
  2007{\natexlab{b}}.

\bibitem[{\textit{Peter}(2009)}]{peter2008crd}
Peter, M., {Coupled reaction--diffusion processes inducing an evolution of the
  microstructure: Analysis and homogenization}, \textit{Nonlinear Analysis},
  \textit{70}, 806--821, 2009.

\bibitem[{\textit{Ricard}(2007)}]{ricard2007pmc}
Ricard, Y., Physics of mantle convection, in \textit{{Treatise on Geophysics}},
  vol.~7, edited by G.~Schubert, Elsevier, 2007.

\bibitem[{\textit{Ricard and Bercovici}(2003)}]{ricard2003tpd}
Ricard, Y., and D.~Bercovici, {Two-phase damage theory and crustal rock
  failure: the theoretical'void'limit, and the prediction of experimental
  data}, \textit{Geophysical Journal International}, \textit{155}(3),
  1057--1064, 2003.

\bibitem[{\textit{Ricard et~al.}(2001)\textit{Ricard, Bercovici, and
  Schubert}}]{ricard2001b}
Ricard, Y., D.~Bercovici, and G.~Schubert, {A two-phase model of compaction and
  damage, 2: Applications to compaction, deformation, and the role of
  interfacial surface tension}, \textit{Journal of Geophysical Research},
  \textit{106}, 8907--8924, 2001.

\bibitem[{\textit{Sanchez-Palencia}(1980)}]{sanchezpalencia1980nhm}
Sanchez-Palencia, E., \textit{{Non-homogeneous media and vibration theory}},
  Lecture Notes in Physics, 127, 1980.

\bibitem[{\textit{Schmeling}(2000)}]{schmeling2000pma}
Schmeling, H., {Partial melting and melt segregation in a convecting mantle},
  in \textit{{Physics and Chemistry of Partially Molten Rocks}}, edited by
  N.~Bagdassarov, D.~Laporte, and A.~Thompson, pp. 141--178, Kluwer Academic,
  2000.

\bibitem[{\textit{Scott and Stevenson}(1984)}]{scott1984ms}
Scott, D., and D.~Stevenson, {Magma solitons}, \textit{Geophysical Research
  Letters}, \textit{11}(11), 1161--1161, 1984.

\bibitem[{\textit{Scott and Stevenson}(1986)}]{scott1986map}
Scott, D., and D.~Stevenson, {Magma ascent by porous flow}, \textit{Journal of
  Geophysical Research}, \textit{91}, 9283--9296, 1986.

\bibitem[{\textit{Simpson}(2008)}]{simpson08t}
Simpson, G., The mathematics of magma migration, Ph.D. thesis, Columbia
  University, 2008.

\bibitem[{\textit{Simpson et~al.}(2008{\natexlab{a}})\textit{Simpson,
  Spiegelman, and Weinstein}}]{simpson08a}
Simpson, G., M.~Spiegelman, and M.~Weinstein, A multiscale model of partial
  melts 1: Effective equations, submitted to Journal of Geophysical Research,
  2008{\natexlab{a}}.

\bibitem[{\textit{Simpson et~al.}(2008{\natexlab{b}})\textit{Simpson,
  Spiegelman, and Weinstein}}]{simpson08b}
Simpson, G., M.~Spiegelman, and M.~Weinstein, A multiscale model of partial
  melts 2: Numerical results, submitted to Journal of Geophysical Research,
  2008{\natexlab{b}}.

\bibitem[{\textit{Spiegelman}(1993)}]{Spiegelman1993a}
Spiegelman, M., Flow in deformable porous media. part 1: Simple analysis,
  \textit{Journal of Fluid Mechanics}, \textit{247}, 17--38, 1993.

\bibitem[{\textit{Spiegelman}(2003)}]{spiegelman2003lam}
Spiegelman, M., {Linear analysis of melt band formation by simple shear},
  \textit{Geochemistry, Geophysics, Geosystems}, \textit{4}(9), 8615, 2003.

\bibitem[{\textit{Spiegelman et~al.}(2007)\textit{Spiegelman, Katz, and
  Simpson}}]{spiegelman07aia}
Spiegelman, M., R.~Katz, and G.~Simpson, {An Introduction and Tutorial to the
  ``McKenize Equations" for magma migration},
  \url{http://www.geodynamics.org/cig/workinggroups/magma/workarea/benchmark/M%
cKenzieIntroBenchmarks.pdf}, 2007.

\bibitem[{\textit{Stevenson and Scott}(1991)}]{stevenson1991mfr}
Stevenson, D., and D.~Scott, {Mechanics of Fluid-Rock Systems}, \textit{Annual
  Review of Fluid Mechanics}, \textit{23}(1), 305--339, 1991.

\bibitem[{\textit{Temam}(2001)}]{temam2001nse}
Temam, R., \textit{{Navier-Stokes Equations: Theory and Numerical Analysis}},
  American Mathematical Society, 2001.

\bibitem[{\textit{Torquato}(2002)}]{torquato2002rhm}
Torquato, S., \textit{{Random Heterogeneous Materials: Microstructure and
  Macroscopic Properties}}, Springer, 2002.

\bibitem[{\textit{Wark and Watson}(1998)}]{wark1998gsp}
Wark, D., and E.~Watson, {Grain-scale permeabilities of texturally
  equilibrated, monomineralic rocks}, \textit{Earth and Planetary Science
  Letters}, \textit{164}(3-4), 591--605, 1998.

\end{thebibliography}

%
%
%
%
%
%
%
%


%
%

\end{article}





%
%
%
%
%
%


\end{document}